\documentclass[fleqn,usenatbib]{mnras}

\usepackage{newtxtext,newtxmath}

\usepackage[T1]{fontenc}

\DeclareRobustCommand{\VAN}[3]{#2}
\let\VANthebibliography\thebibliography
\def\thebibliography{\DeclareRobustCommand{\VAN}[3]{##3}\VANthebibliography}

\usepackage{graphicx}
\usepackage{float}
\usepackage{amsmath}	

\usepackage{threeparttable}
\graphicspath{{img/}}
\usepackage[normalem]{ulem}

\usepackage{xcolor}

\newcommand{\cm}{cm$^{-1}$}

\newcommand{\exocross}{\textsc{ExoCross}}
\newcommand{\molpro}{\textsc{MOLPRO}}

\newcommand{\duo}{{\sc Duo}}
\newcommand{\Duo}{{\sc Duo}}

\newcommand{\ai}{\textit{ab initio}}

\newcommand{\X}{\mbox{${X}\,^{2}\Delta$}}
\newcommand{\W}{\mbox{${W}\,^{2}\Pi$}}
\newcommand{\V}{\mbox{${V}\,^{2}\Sigma^{+}$}}

\newcommand{\name}{BYOT}

\title[ExoMol line lists -- {LXV}. NiH]{ExoMol line lists -- LXV. Mid-Infrared rovibronic spectroscopy of  isotopologues of NiH}

\author[]{
Kirill Batrakov,$^1$
Sergei N. Yurchenko,$^{1}$
Alec Owens,$^{1}$
Jonathan Tennyson,$^{1}$\thanks{The corresponding author: j.tennyson@ucl.ac.uk}
\newauthor{Alexander Mitrushchenkov,$^{2}$
Amanda J. Ross,$^{3}$ Patrick Crozet,$^{3}$ Asen Pashov$^{4}$}
\vspace*{4mm}\
\\
$^1$ Department of Physics and Astronomy, University College London, Gower Street, WC1E 6BT London, UK \\
$^2$ MSME, Universit\'{e} Gustave Eiffel, CNRS UMR 8208, Univ Paris Est Creteil, F-77474 Marne-la- Vall\'{e}e, France\\
$^{3}$University of Lyon, Universit\'{e} Claude Bernard Lyon 1 \& CNRS, Institute Lumi\'{e}re Mati\'{e}re UMR 5309, F-69622, Villeurbanne, France\\
$^4$Faculty of Physics, Sofia University, 5 James Bourchier Boulevard, 1164 Sofia, Bulgaria
}

\date{Accepted XXXX. Received XXXX; in original form XXXX}

\date{\today}

\begin{document}

\label{firstpage}

\maketitle

\pagerange{\pageref{firstpage}--\pageref{lastpage}}

\begin{abstract}
 New line lists for four isotopologues of nickel monohydride, $^{58}$NiH, $^{60}$NiH, $^{62}$NiH, and $^{58}$NiD are presented covering the wavenumber range $<10000$ cm$^{-1}$ ($\lambda > 1$~$\mu$m), $J$ up to 37.5 for transitions within and between the three lowest-lying electronic states, \X, \W, and \V. The line lists are applicable for temperatures up to 5000 K. The line lists calculations are based on a recent empirical NiH spectroscopic model  [Havalyova et al. \textit{J. Quant. Spectrosc. Radiat. Transf.}, \textbf{272}, 107800, (2021)] which is adapted for the variational nuclear-motion code \textsc{Duo}. The model consists of  potential energy curves, spin-orbit coupling curves, electronic angular momentum curves, spin-rotation coupling curves, $\Lambda$-doubling correction curve for $^2\Pi$ states and Born-Oppenheimer breakdown (BOB) rotational correction curves. New \textit{ab initio} dipole moment curves, scaled to match the experimental dipole moment of the ground state, are used to compute Einstein A coefficients. The \name\ line lists are included in the ExoMol database at \url{www.exomol.com}.

\end{abstract}

\begin{keywords}
line: profiles - molecular data - exoplanets - stars: atmospheres - stars: low-mass
\end{keywords}



\section{Introduction}

Ni has  a cosmic  abundance similar to that of chromium or calcium, so the nickel hydride (NiH)
molecule could reasonably be considered a target species for astrophysical observations. Its
spectrum in the visible region happens to be overlapped by strong bands of TiO, possibly explaining why
it has not joined CrH and FeH in the list of well-known absorbers in stellar atmospheres, but
strong rotation-vibration and ro-vibronic systems are to be expected further to the infrared.
This paper focuses on transitions occurring within the three low-lying 'supermultiplet' states
of NiH,  characterized by large permanent dipole moments close to their equilibrium
internuclear distances.  Few of these transitions have been observed directly, but many can be
predicted from term energies extracted from potential energy curves and coupling functions
derived from the analysis of optical data.

We have adjusted the model proposed by  \citet{21HaBoPa.NiH} to suit the framework of program \duo\ \citep{Duo}, and have created an NiH line list as part of the ongoing ExoMol project \citep{jt528,jt939}. This entailed computation of permanent and transition dipole moments, as well energy levels, and then compilation of an optimized data base where experimentally-derived energies replace calculated quantities wherever possible.
These line lists should aid observation and assignments of NiH  in cool stars and elsewhere.


\begin{figure*}
\includegraphics[width=0.45\textwidth]{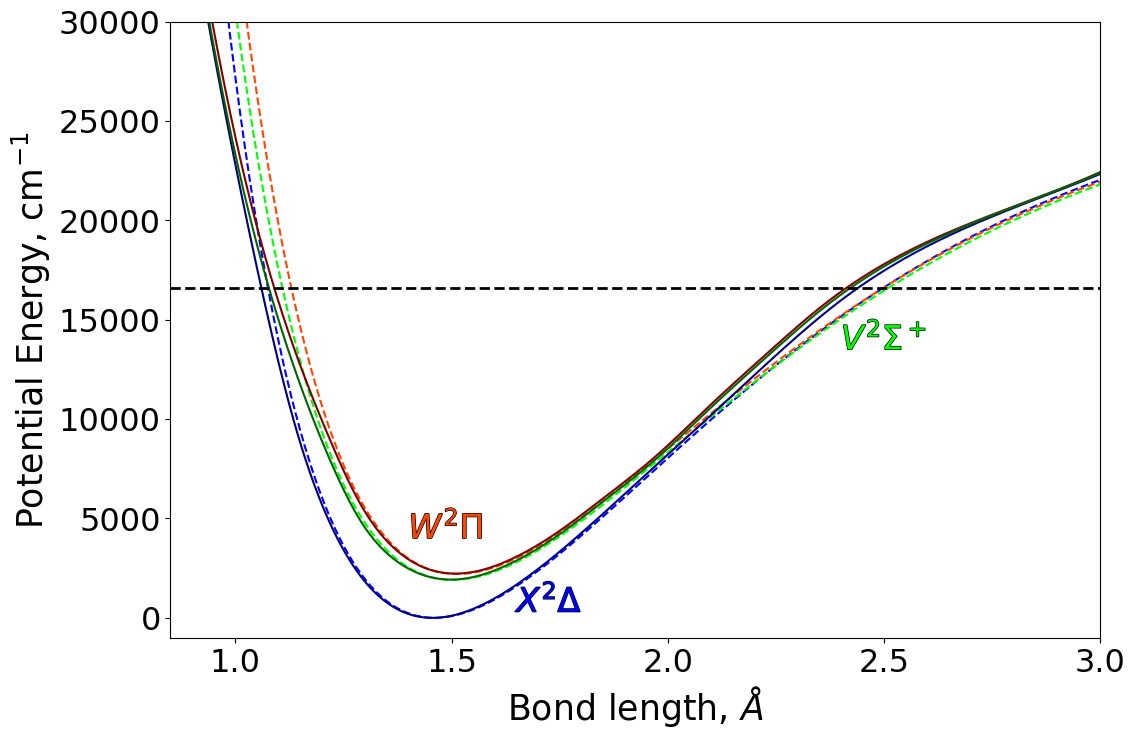}
\includegraphics[width=0.45\textwidth]{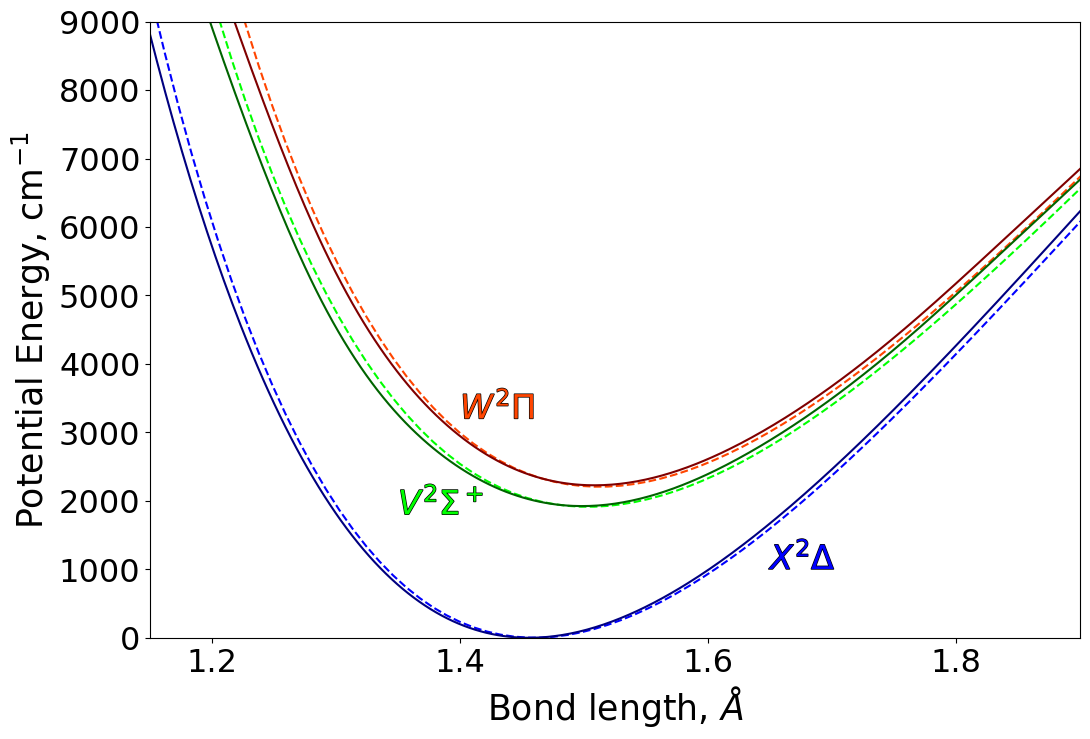}
   \label{f:PEC}
\caption{ Comparison of our refined PECs of NiH to those from \citet{21HaBoPa.NiH}. Darker shades of colours correspond to the curves from \citet{21HaBoPa.NiH}. Lighter shades of colours correspond to our refined curves. The horizontal dashed line shows where the first of higher-lying electronic states with allowed transitions start according to \citet{07ZoLixx.NiH}. 
The right display zooms in on the spectroscopically important region.} 
 \end{figure*}

Brown and Evenson and their co-workers \citep{88BeEvNe.NiH,91NeBeEv.NiH,93BrBeEv.NiH}
were amongst the first to provide spectroscopic data on NiH with the specific purpose of supporting the search for NiH in astrophysical media.
They used laser magnetic resonance to measure pure rotation in the far IR.  Around the same time, the Urban group explored transitions within the 2D ground state manifold, and between the  \X\ and \W\ states of NiH and NiD \citep{89LiSiBa.NiH,93LiUrEv.NiH,91BaUrNe.NiH},
establishing energy levels more accurately than the earlier emission studies from a hollow cathode lamp  \citep{82ScLoKa.NiH} or absorption \citep{87KaLoSc.NiH} in a King furnace had done.
\citet{90StNaSh.NiH} measured $\Lambda$-doubling splittings in microwave spectra of the ground state of $^{58}$NiH and $^{60}$NiH, noting the influence of mixing between states, since `normal' $\Delta$  states were not then expected to show significant $\Lambda$-doubling.
In later years, laser excitation and fluorescence in the optical domain became increasingly important in the study of NiH, with work notably from the Field group at MIT providing a great deal of insight into the electronic structure of NiH.
\citet{90HiFixx.NiH} observed the  \W\ and \V\ states through laser-induced fluorescence, giving further details in \citet{90KaScGr.NiH}. Their supermultiplet model \citep{90GrLiFi.NiH} was published shortly afterwards, demonstrating that atomic parameters can account for much of the molecular structure in these three low-lying states, including effective Land\'e factors determined from Zeeman spectroscopy \citep{89LiFiel.NiH,97McKaSt.NiH}.
 Intracavity laser absorption spectroscopy approach was used by \citet{05OBOB.NiH, 08ShNuBr.NiH, 09ShSoLi.NiH} to study  $B\,^2 \Delta$ -- \X\  transitions for the four dominant nickel isotopes. All these studies facilitated work from the Lyon group, who recorded isotopically selective dispersed fluorescence spectra on a Fourier transform instrument, both for NiH \citep{12RoCrRi.NiH,09VaRiCr.NiH} and NiD \citep{18AbShCr.NiH,19RoCrAd.NiH}, using a sputtering source operating around 450 K.


Theoretical studies of NiH started with \citet{77GuBlKu.NiH} who made an early attempt to describe the potential energy curves (PECs) of NiH. However, this study was not successful in describing the complex electronic structure of the molecule, as they did not show that PECs corresponding to the three lowest electronic states \X, \W, \V\ have similar shapes and lie close to each other. Later, \citet{81BaBjxx.NiH} identified this behaviour which was confirmed by \citet{82BlSiRo.NiH}. Further development of theoretical methods for calculations of NiH PECs and energy levels were carried out by \citet{83WaBaxx.NiH}, \citet{84RuBlHe.NiH}. A more precise description of relativistic effects in the molecule was obtained by \citet{89MaBlSi.NiH}. Due to the similarity of PECs of the three lowest electronic states, a supermultiplet approach, which does the fitting of energy levels for all electronic states simultaneously, was proposed in \citet{91GrLiNe.NiH}. Further theoretical work was carried out by \citet{94PoMeNe.NiH} with a focus on chemical bonds of NiH, by \citet{04DiChDo.NiH} with a focus on the application of hybrid density functional theory and by \citet{08GoeMas.CrH} who compared different techniques on 3d transition metal hydrides. \citet{07ZoLixx.NiH} presented probably one of the most detailed \ai\ studies on NiH PECs, as they considered several higher-lying electronic states as well.

Studies on the NiH dipole moments are less extensive. The first experimental study of electric dipole moment properties of NiH was carried out by \citet{85GrRiFi.NiH} who measured the ground state dipole moment,  $2.4 \pm 0.1$ Debye. Later, \citet{08ChStxx.NiH} confirmed and refined the previous result to $2.44 \pm 0.02$ Debye. Different theoretical methods were used to calculate dipole moments by \citet{85WaBaLa.NiH}, who computed values in the range $1.74-3.98$ Debye using different techniques, and \citet{90BaLaKo.NiH}, who computed values in the range $1.544-3.310$ Debye using different techniques. Thus the agreement between experiments and theoretical techniques remain uncertain.

\citet{21HaBoPa.NiH} presented a spectroscopic model for low-lying electronic states of NiH. This model accurately  predicts the energy term values for the three lowest-lying electronic states of $^{58}$NiH, $^{60}$NiH, $^{62}$NiH isotopologues. \citet{21HaBoPa.NiH} fitted cubic spline pointwise (PECs and relevant couplings) curves to experimentally derived energies of NiH from a series of spectroscopic studies by  \citet{90StNaSh.NiH,92LiBaUr.NiH,09VaRiCr.NiH,12RoCrRi.NiH,18AbShCr.NiH}.
However, \citet{21HaBoPa.NiH} made no mention of transition intensities.  \citet{18AbShCr.NiH} presents the most detailed combined experimental-theoretical  study of the $^{58}$NiD isotopologue up to date, giving the model in terms of Dunham-type parameters, again with no  information on the intensities. Since that paper was
published, more low-$\Omega$ levels of NiD have been reported by \citet{19RoCrAd.NiH}, motivating further analysis.

In this work, we  provide comprehensive infrared ($\lambda > 1$ $\mu$m) line lists for four isotopologues  of nickel hydride ($^{58}$NiH, $^{60}$NiH, $^{62}$NiH, and $^{58}$NiD) in their  three lowest electronic states (\X, \W, \V), containing transition frequencies and Einstein A coefficients, as well as data on individual energy levels, including their lifetimes. To this end, we combine the  spectroscopic model from \citet{21HaBoPa.NiH} with a new set of \ai\ dipole moment curves (DMCs). The line lists were computed using the \Duo\ software \citep{Duo}.\footnote{\Duo\ is a Fortran 2003 program freely available from \url{www.github.org/exomol}} The potential energy and coupling curves of \citet{21HaBoPa.NiH} are refitted to make them compatible with \Duo, while the \ai\ DMCs are scaled to the experiment \citep{08ChStxx.NiH}.   We believe that these line lists are robust enough to be used in astrophysical observations of spectra and the detection of NiH in astrophysical media.

\section{Spectroscopic Model}

\subsection{Energy calculations}

We used the \Duo\ \citep{Duo} program to solve the coupled system of Schr\"{o}dinger equations. The grid of 301 points with the bond length range $0.75-4$ \AA\ and a basis set of $v_{max} = 40$ were used in conjunction with the Sinc DVR method. The details of the \Duo\ methodology can be found in \citet{Duo}.

As a starting point in our calculations we used the empirical spectroscopic model of NiH recently produced by \citet{21HaBoPa.NiH}. Their model consists of three PECs, all relevant spin-orbit curves (SOCs), electronic angular momentum curves (EAMCs, also known as $L$-uncoupling curves), spin-rotation curves (SRCs)  as well as  Born-Oppenheimer breakdown (BOB) curves. The curves are given in a pointwise grid representation in conjunction with cubic spline interpolation. Due to  functional representations of the off-diagonal couplings   between \Duo\ and the computational method used by \citet{21HaBoPa.NiH}, we could not port their model  directly into  \Duo\ one-to-one. For example, \Duo\ uses a different representation for the BOB and  $\Lambda$-doubling  effects. Hence, we could not reproduce their results with \Duo\ with sufficiently good accuracy and decided to further transform  their model by refitting to the same set of the experimental data, covering  experimentally derived energies of $^{58}$NiH, $^{60}$NiH and $^{62}$NiH
in their   three lowest electronic states \X, \W, \V\ as reported in \citet{21HaBoPa.NiH}.  We also include the $^{58}$NiD  isotopologue into the current work and build and a similar empirical model  for which experimental energy term values from \citet{18AbShCr.NiH} were used.

As far as the intensity calculations are concerned, we have computed electric (transition) dipole moment curves of NiH between the electronic states \X, \W, \V\ \ai\ as discussed below.

\subsection{Analytical Description of the Spectroscopic Model of NiH}

In order to make the spectroscopic model by \citet{21HaBoPa.NiH} compatible with \Duo, we had to convert their curves from their pointwise grid representation to an analytic representation used in \Duo.
In the following, we introduce the analytic representation of the curves used in the refinement of the spectroscopic model by \citet{18AbShCr.NiH} and discuss the effect of this modifications.

An Extended Morse Oscillator (EMO) function \citep{EMO} was used to represent PECs for all electronic states:
\begin{equation}
    V(r) = V_{\rm e} + (A_{\rm e}-V_{\rm e}) \left[ 1 - \exp{ \left( - \sum^N_{k=0} B_k \xi^k_p (r-r_{\rm e}) \right) } \right],
    \label{eq:EMO}
\end{equation}
where $r_{\rm e}$ is an equilibrium distance, $A_{\rm e}$ is a dissociation asymptote, $A_{\rm e} - V_{\rm e}$ is a dissociation energy, $\xi_p$ is the \v{S}urkus variable given by
\begin{equation}
    \xi_p = { \frac{r^p - r_{\rm e}^p}{r^p + r_{\rm e}^p} }.
    \label{eq:surkus}
\end{equation}
where $p$ the exponent is chosen as part of the fit.

\begin{figure}
\includegraphics[width=0.45\textwidth]{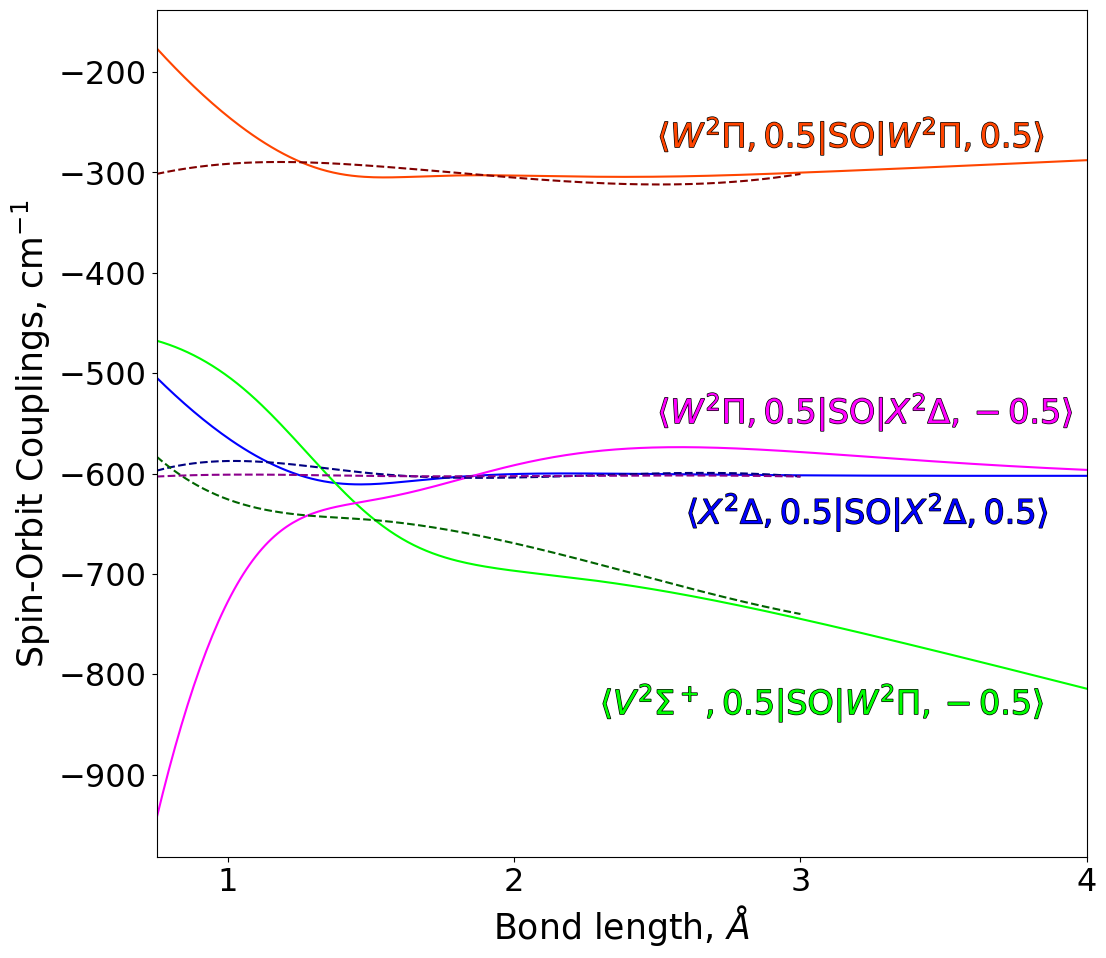}
\caption{ SOC obtained in this work (solid lines and lighter shades of colours) and by \citet{21HaBoPa.NiH} (dashed lines and darker shades of colours). Captions on the figure show what are the electronic states involved and corresponding values of   $\Sigma$, the projection of the spin of electrons on the molecular axis. }
    \label{f:SO_comparison}
\end{figure}

Figure \ref{f:PEC} shows the original PECs by \citet{21HaBoPa.NiH} and compares them to our refined PECs. The horizontal dashed line at 16500~\cm\ indicates the region of the strong spectroscopic presence of NiH in the visible region due to the next family of the doublet electronic states of NiH, see the \ai\ study by \citet{07ZoLixx.NiH}.  The differences caused by the fits are small, especially in the spectroscopically relevant region below 10000~\cm. This is reassuring as essentially we attempt to reconstruct the same spectroscopic model and the differences should be mainly due to different responses of our models to the ambiguities  from the limited amount  of the underlying experimental information. Figure ~\ref{f:SO_comparison} provides a similar comparison between the SOCs,  obtained in this work and  from \citet{21HaBoPa.NiH}. These differences, as well as analogous difference between other couplings from these two studies, is  manifestation of  the strong correlation between different components defining our models and thus is reflection of the complexity of the system.

The following function was initially used to describe all the coupling and correction curves:
\begin{equation}
    F(r) = \sum^N_{k=0} B_k z^k (1-\xi_p)+\xi_p B_{\infty}
    \label{eq:polynom_decay}
\end{equation}
where $z$ is a coordinate damped at large $r$ as given by
\begin{equation}
    z = (r-r_{\text{ref}}) \exp{ \left( -\beta_2 (r-r_{\text{ref}})^2 - \beta_4 (r-r_{\text{ref}})^4 \right) },
    \label{eq:damped_z}
\end{equation}
see \citet{jt703} and \citet{jt711} for more details. Here $r_{\text{ref}}$ is a reference position which equals $r_{\rm e}$ by default, $\beta_2$ and $\beta_4$ are damping parameters, $B_{\infty}$ is a long-range asymptote, usually fixed to 0 or 1.
The functional formed used do not provide any control over short values of $r$ and are theretofore prone to  divergence as a result of the refinement due to the lack of constraints from experimental data at high very energies specific for the short distances.

\begin{figure*}
\centering
\begin{tabular}{cc}
  \includegraphics[width=70mm]{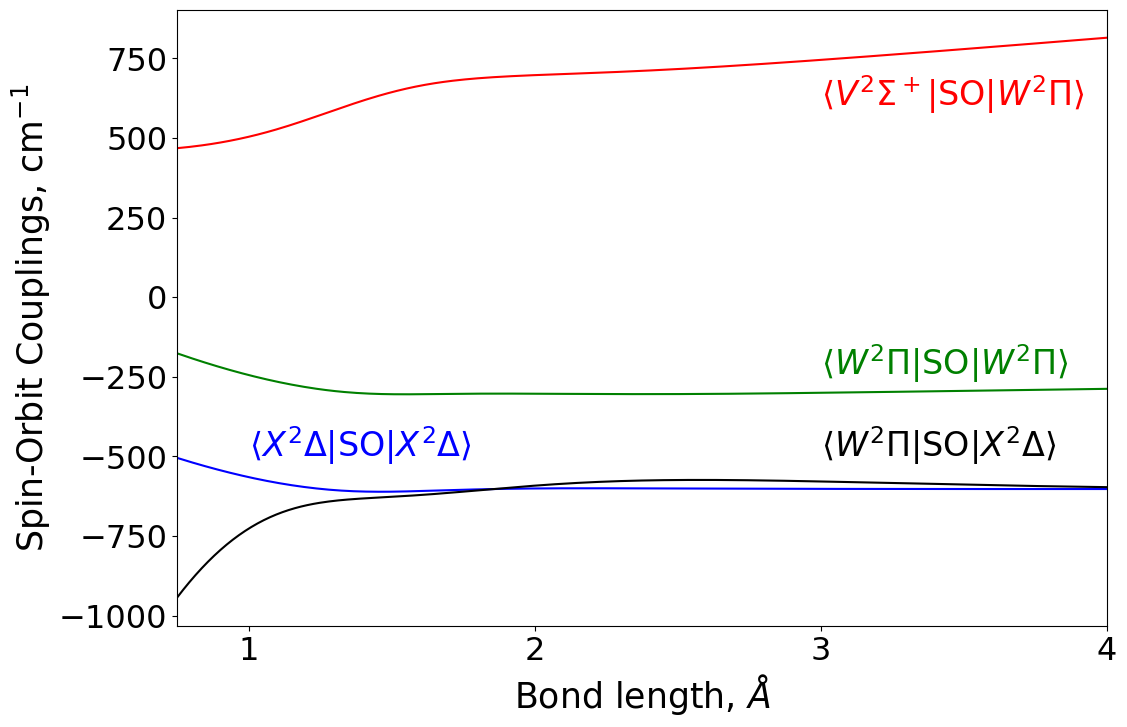} & \includegraphics[width=70mm]{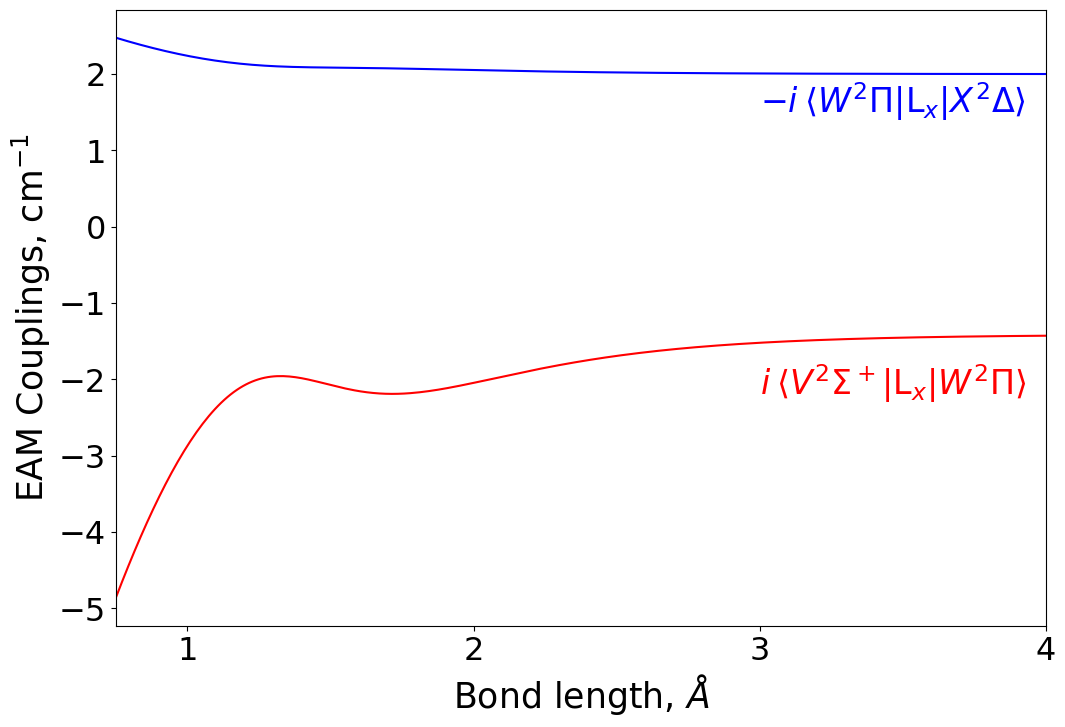} \\
  \includegraphics[width=70mm]{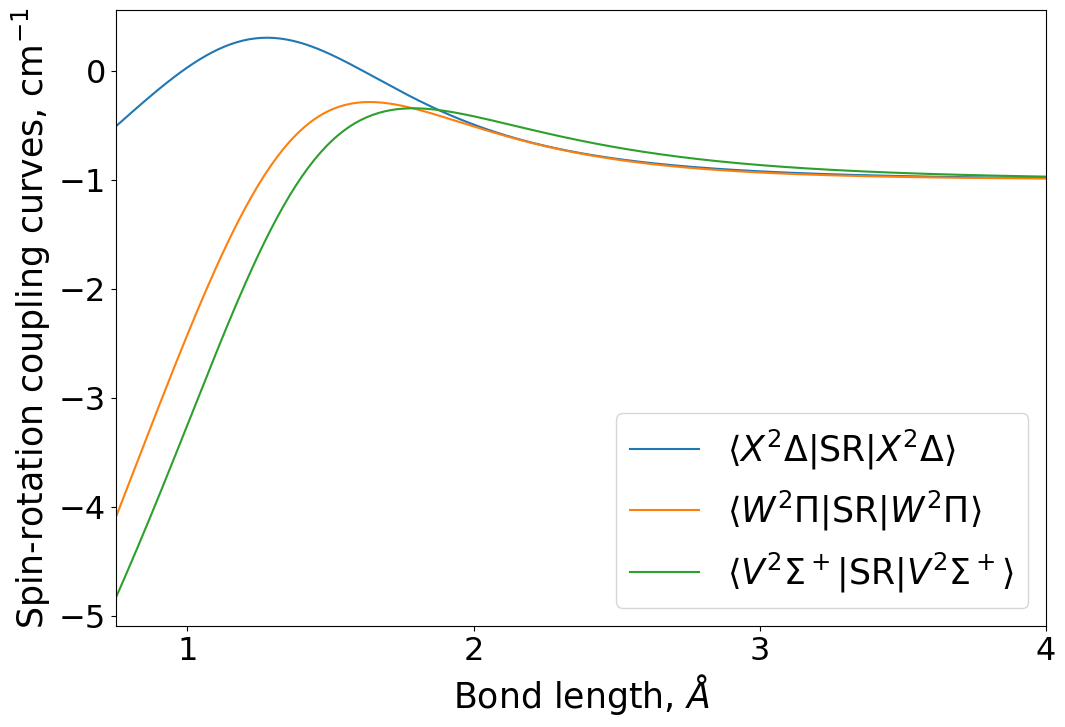} & \includegraphics[width=70mm]{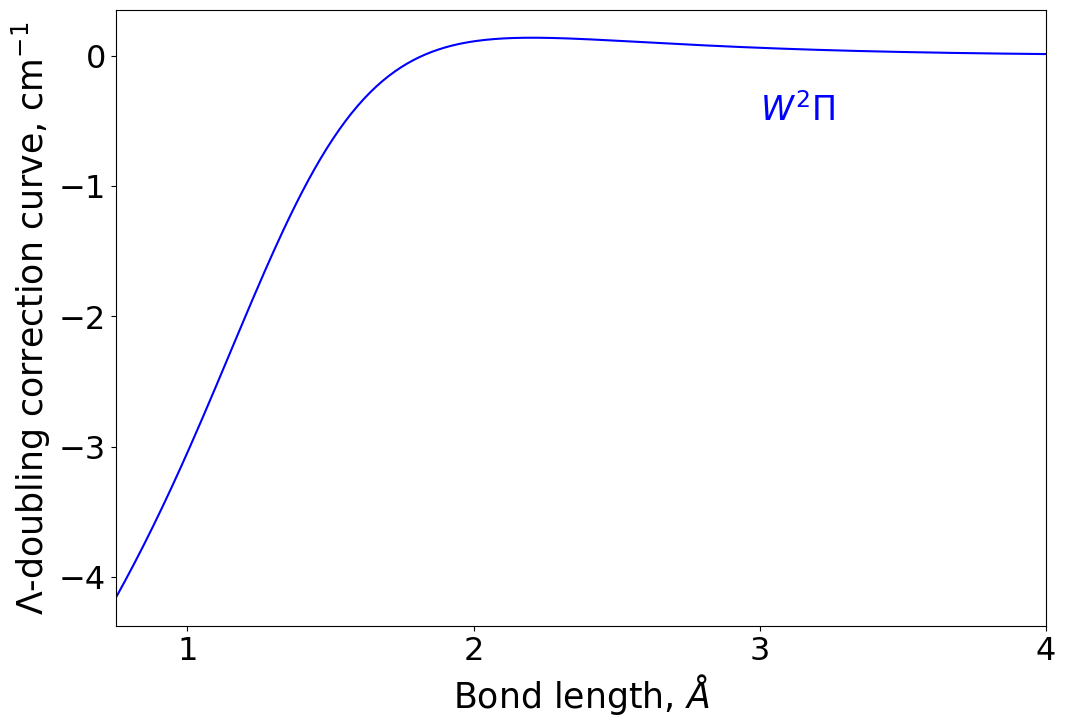} \\
  \includegraphics[width=70mm]{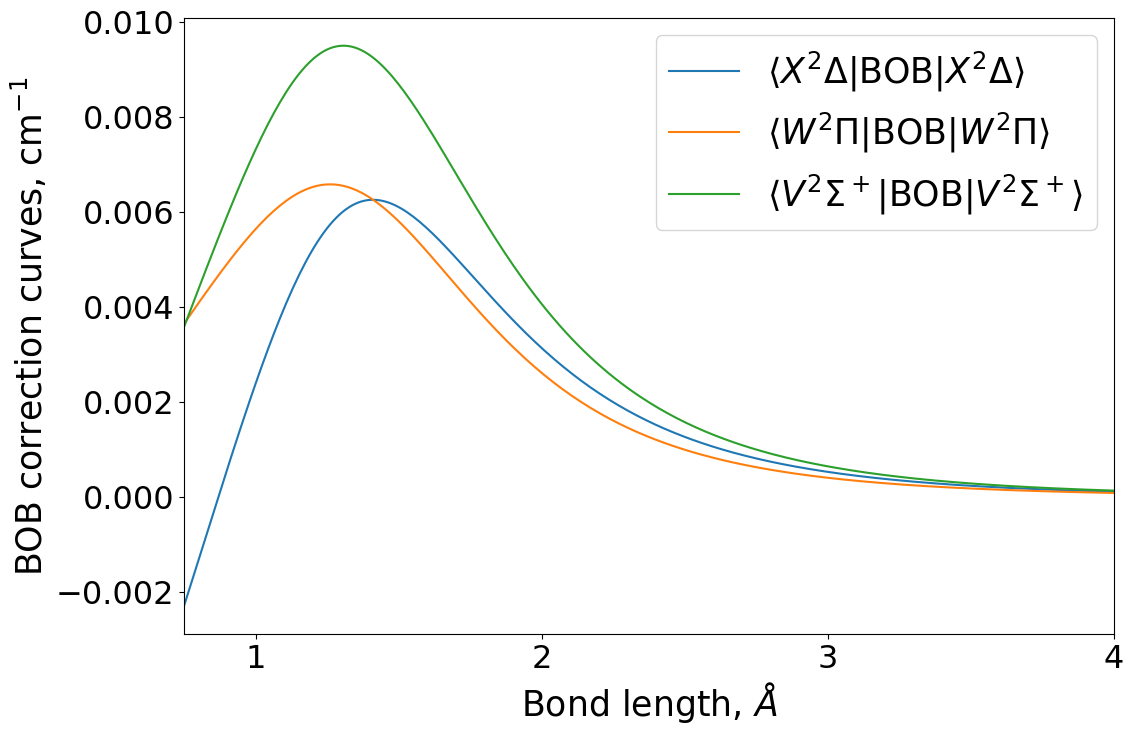} & \includegraphics[width=70mm]{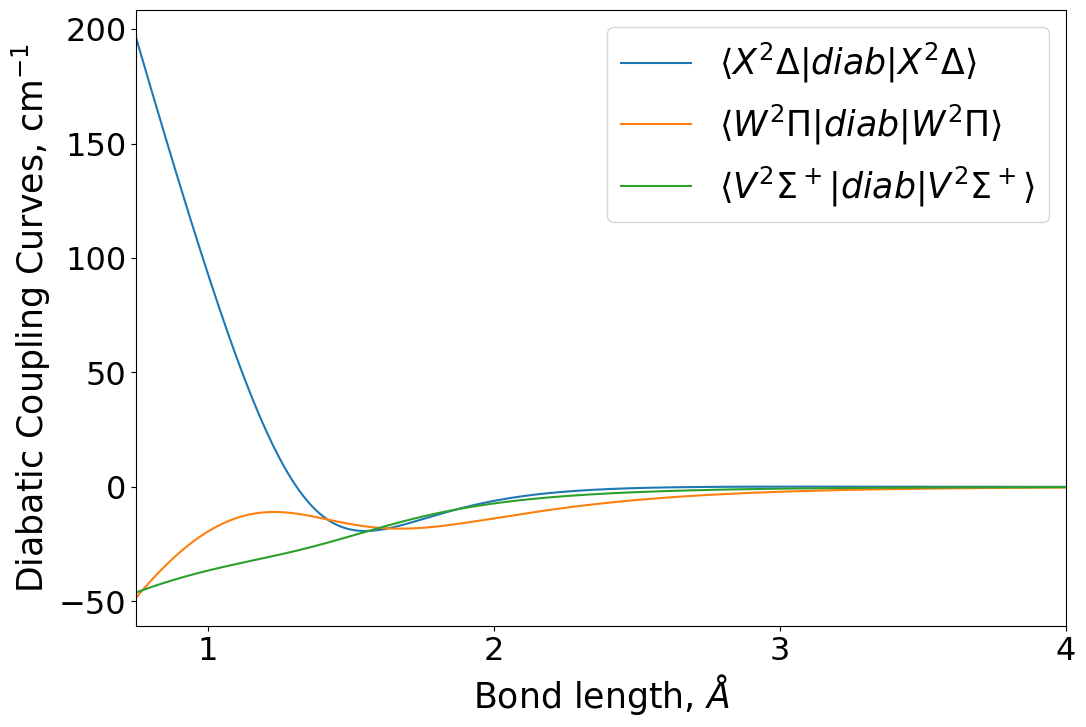} \\
\end{tabular}
\caption{Empirical coupling and correction curves for $^{58}$NiH and diabatic coupling curves for $^{58}$NiD.}
\label{f:couplings}
\end{figure*}

We therefore modified Eq.~\eqref{eq:damped_z} by incorporating a short-distance damping function based on a form proposed by \citet{82DoScMa.ai} as given by
\begin{equation}
    F(r) = D^{\rm DS}(r) \sum^N_{k=0} B_k z^k (1-\xi_p)+\xi_p B_{\infty}
    \label{eq:polynom_decay_damp},
\end{equation}
which differs from the Eq.~\eqref{eq:polynom_decay} by an additional damping term
\begin{equation}
    D^{\rm DS}(r) = \left[ 1 - \exp{ \left( -br - cr^2 \right) } \right].
    \label{eq:Douketis_damp}
\end{equation}
Here, $b$, $c$, and $s$ are short-range damping parameters,  see \citet{82DoScMa.ai,11LeHaTa.MLR,LEVEL} for more details. This damped formula was used for one SOC, $\langle$\W |SO|\X\ $\rangle$, all spin-rotation couplings, and all BOB-rotation corrections. Other couplings and dipole moment curves were modelled using Eq.~\eqref{eq:polynom_decay}. Figure~\ref{f:couplings} shows the final curves for the $^{58}$NiH isotopologue. We used the same curves for $^{60}$NiH and $^{62}$NiH with the exception of the BOB-rotation correction, which were refitted due to the changed mass.  A similar approach was used for the $^{58}$NiD isotopologue, but the quality of the fit decreased by two orders of magnitude, mainly due to more limited experimental coverage. Thus, for $^{58}$NiD we also had to refit diagonal spin-orbit coupling curves ($\langle$\X\ |SO|\X\ $\rangle$ and $\langle$\W |SO|\W $\rangle$) and add diabatic diagonal couplings for \X\ and \W\ states as a correction to their PECs (it turned out to be more efficient than varying PECs themselves). The latter were modelled using Eq.~\eqref{eq:polynom_decay}. The bottom-right display of Fig.~\ref{f:couplings} shows these diabatic coupling curves of $^{58}$NiD.

The root mean square (rms) deviation of energy terms values computed using our model from the experimental values were: $0.036$ \cm\ for $^{58}$NiH, $0.029$ \cm\ for $^{60}$NiH, $0.017$ \cm\ for $^{62}$NiH, and $0.061$ \cm\ for $^{58}$NiD.
$^{60}$NiH and $^{62}$NiH had fewer experimental term values ($J_{\rm max} = 12.5$ and 9.5, respectively) than $^{58}$NiH ($J_{\rm max} = 16.5$). For $^{58}$NiD, $J_{\rm max} = 17.5$.
It is important to note that the model presented by \citet{21HaBoPa.NiH} is more accurate than our work, as they got rms deviations in the range of 0.011 -- 0.015 \cm, which can be  attributed to a more efficient fitting procedure used by \citet{21HaBoPa.NiH}.

\subsection{Dipole Moments}

\begin{figure}
\includegraphics[width=0.45\textwidth]{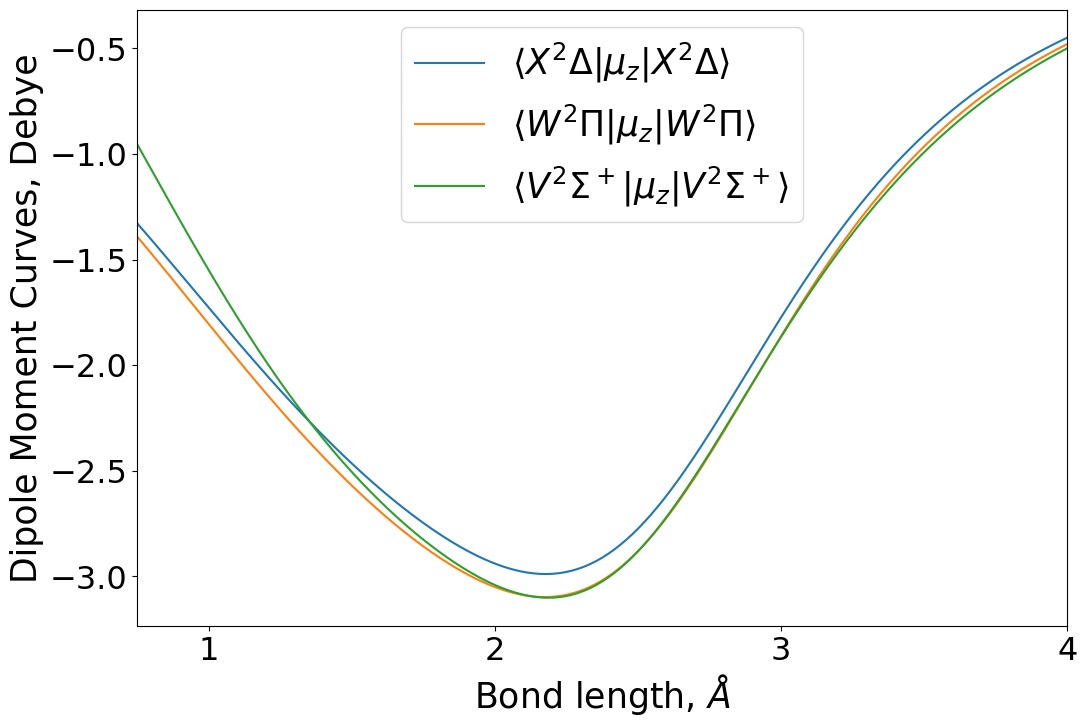}
\caption{ Diagonal dipole moments curves of NiH along the $z$-axis. }
    \label{f:DMZ}
\end{figure}

We used the MCSCF/aug-cc-pVTZ-DK level of theory to compute (transition) dipole moment curves between the three lowest electronic states of NiH using the \molpro\ software \citep{MOLPRO}. The diagonal ones are shown in Fig.~\ref{f:DMZ}. The non-diagonal (transition) dipoles were two orders of magnitude smaller than the diagonal ones ($<0.09$~Debye).


The study by \citet{08ChStxx.NiH} measured the permanent dipole moment for the ground \X\ state using the optical Stark effect, for which they obtained 2.44 Debye. This value can be compared to the  value of the permanent moment of \X\ state of 3.13~D at $r=r_{\rm e}$ and to the vibrationally averaged value ($v=0$) of 3.16~D computed with \Duo. Increasing the \ai\ level of theory to significantly more expensive MRCI/aug-cc-pVTZ-DK only reduced it to 3.1~D, i.e had a negligible effect. We therefore decided to empirically adjust the DMC of \X\ by  scaling our MCSCF curve by the factor of the ratio of the experiment over theory  ($2.44/3.16 = 0.77$). Unfortunately, there is no experimental information on other two diagonal dipole moments, \W\ and \V. Considering that all three \ai\ curves appear to closely follow each other, see Fig. \ref{f:DMZ}, we decided to apply the same scaling of 0.77 also to these two curves. The non-diagonal (transitional) dipoles were not scaled, as their shapes do not follow the diagonal curves. Dipole moment curves were not changed between different isotopologues.

The three diagonal DMCs were then modelled using the analytic description of Eq.~\eqref{eq:polynom_decay}. This was important to reduce numerical noise typical for high vibrational overtones leading to the so-called ``overtone plateaus'' \citep{16MeMeSt}. As part of this treatment, we also applied a threshold to vibrational transitional moments of $10^{-8}$~Debye.

Spectroscopic models for all isotopologues can be found in the supplementary data in the form of input files for \Duo.

\section{Line list}
Line lists for four isotopologues of NiH, $^{58}$NiH, $^{60}$NiH, $^{62}$NiH, and $^{58}$NiD were generated
using the empirical spectroscopic model constructed and the scaled \ai\ MCSCF  DMCs for \X, \W, and \V\ states. The line lists \name\ cover the wavenumber range up to 10000~\cm\ with rotation excitations up to $J=56.5$ (NiH) and $J=78.5$ (NiD). We used the lower state cutoff of  20000 \cm\ aiming at the completeness at high temperatures even at lower frequencies.  Table \ref{t:ll_stats} shows the number of states and transitions computed for each isotopologue.

The line lists are provided in the standard ExoMol format \citep{jt548,jt810} with  States and Trans files. Tables \ref{t:states} and \ref{t:trans} show examples of extracts from these files for $^{58}$NiH. The States file contains the following information for each energy term: counting numbers as state IDs, energies (\cm, either experimental or calculated) and their uncertainties (\cm), lifetimes (s), and standard set of quantum numbers generated by \duo. The Trans file contains state counting numbers (IDs) for upper and lower states, Einstein A coefficients (s$^{-1}$), and transition wavenumbers (\cm).

\begin{table}
    \centering
    \caption{Statistics on the line lists computed.}
      \label{t:ll_stats}
    \begin{tabular}{c|c|r}
    \hline
    \hline
         Isotopologue & No. of states & No. of transitions \\
         \hline
         $^{58}$NiH & 11976 & 643809 \\
         $^{60}$NiH & 11979 & 643851 \\
         $^{62}$NiH & 11985 & 644512 \\
         $^{58}$NiD & 23187 & 1716206 \\
    \hline
    \hline
    \end{tabular}
\end{table}

\begin{figure}
\includegraphics[width=0.45\textwidth]{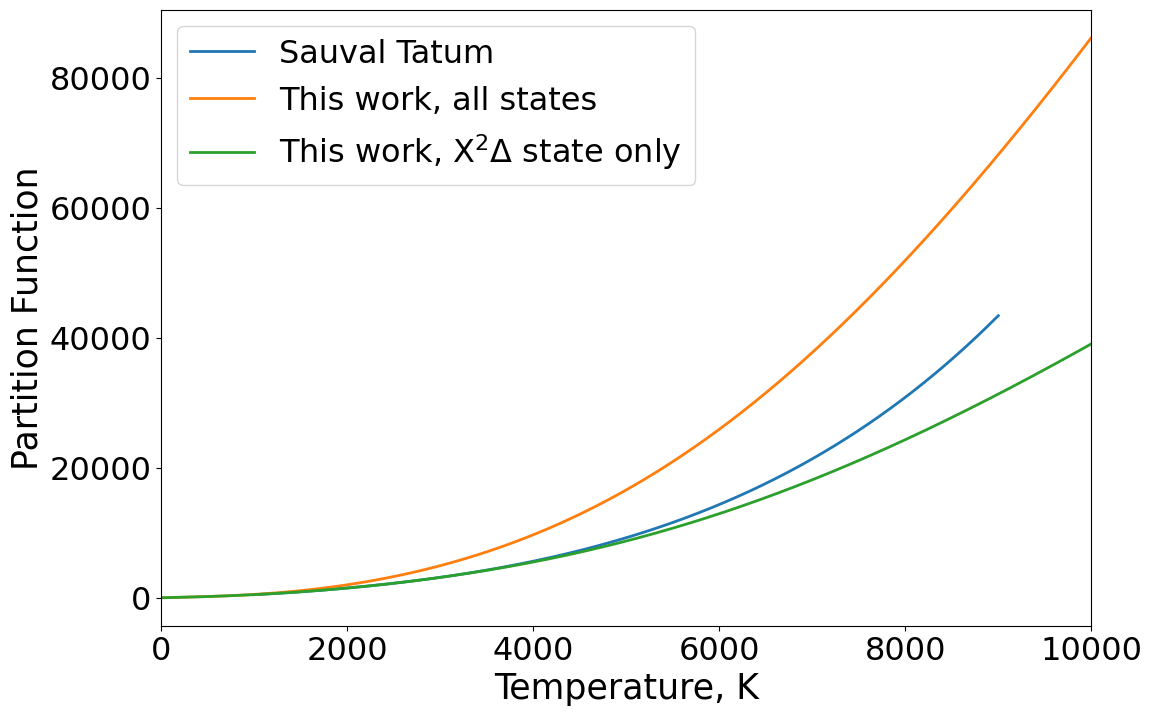}
\caption{Our partition function of $^{58}$NiH (upper curve) compared to that of \citet{84SaTaxx.partfunc} (middle curve). The lower curve is also our partition function but without the \W\ and \V\ energy contributions (see text for details).
}
    \label{f:pf}
\end{figure}

The partition functions for all four isotopologues of NiH were generated using \exocross\ \citep{ExoCross} by summing over the energy levels up to $J = 60.5$ (separate States files were compiled for this). The $^{58}$NiH values were compared to estimates given by \citet{84SaTaxx.partfunc} as shown in Fig.~\ref{f:pf} for temperature up to 10000 K. For this comparison, a  factor of 2 was applied to the values of \citet{84SaTaxx.partfunc} to account for the nuclear spin degeneracy of $^{58}$NiH. This is because ExoMol's convention is to include the  nuclear spin degeneracy into ExoMol partition functions \citep{19PaYuTe}. There is a large discrepancy between us and \citet{84SaTaxx.partfunc}, with our values being significantly larger indicating possible under-count of states in \citet{84SaTaxx.partfunc}, despite  \citet{84SaTaxx.partfunc}'s claim that their partition function's estimate should be valid for temperatures in the range of 1000-9000 K. According to \citet{84SaTaxx.partfunc}, their partition functions are based on the spectroscopic parameters from \citet{79HuHe.book}, which does not appear to cover the \W\ and \V\ electronic states. As a test of the assumption that the difference is due to the lack of the \W\ and \V\ energies, in Fig.~\ref{f:pf} we show a partition function of $^{58}$NiH produced using  the  \X\ state only  (i.e. excluding \W\ and \V). This time the agreement with \citet{84SaTaxx.partfunc}  is very close up to about  6000~K, which supports our assumption.  We conclude that (i) the partition function of NiH by \citet{84SaTaxx.partfunc} is severally incomplete even at moderate temperatures and should not be used and (ii) our partition function should be complete at least up to 6000~K if we believe completeness of \X\ of \citet{84SaTaxx.partfunc}.  We note that the more recent NiH partition function of \citet{16BaCoxx.partfunc} is based on use of the same
data as \citet{84SaTaxx.partfunc} and gives similarly underestimated results.

The completeness of our line list is also affected by the energy threshold used (20000~\cm\ for the lower state), and even more so by the absence of the upper electronic states in our model. The next doublet electronic state of NiH lies at about 16600~\cm\ \citep{07ZoLixx.NiH}, which overlaps with the energy coverage in our line list.

\begin{figure}
\includegraphics[width=0.45\textwidth]{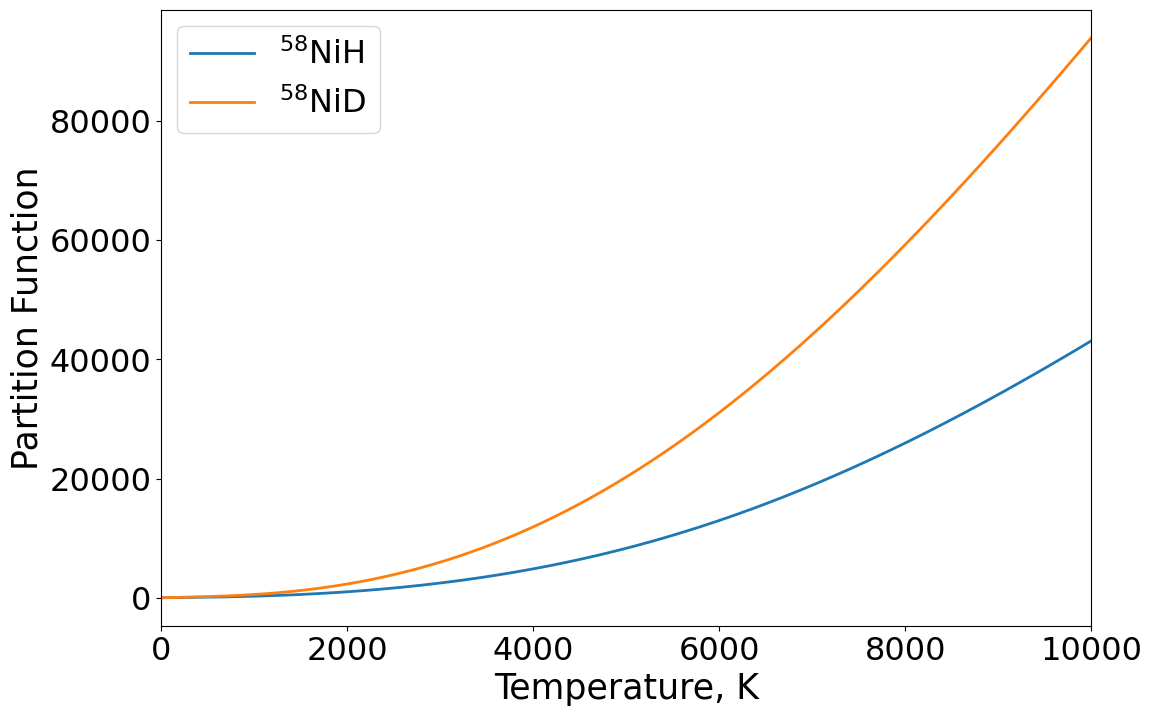}
\caption{Comparison of our partition function of $^{58}$NiD (upper curve) and $^{58}$NiH (lower curve) divided by their respective nuclear spin multiplicity factors.
}
    \label{f:pf:DH}
\end{figure}

Figure \ref{f:pf:DH} shows the comparison of partition functions for $^{58}$NiH and $^{58}$NiD computed with \Duo\ and \exocross. These partition functions were divided by the respective nuclear spin multiplicity factors (2 for $^{58}$NiH and 3 for $^{58}$NiD), so that the only difference between these partition functions is due to different number of states.  $^{58}$NiD contains more states within the same temperature range compared to $^{58}$NiH, see Table~\ref{t:ll_stats}, and thus leads to a larger value of the partition function.


\begin{table*}
\centering
\caption{ Extract from the states file of the line list for  $^{58}$NiH.} 
\label{t:states}
{\tt  \begin{tabular}{rrrrrcclrrrrrrr} \hline \hline
$i$ & $\tilde{E}$ (\cm) & $g_i$ & $J$ & unc. (\cm) & $\tau$ (s$^{-1}$) & \multicolumn{2}{c}{Parity} 	& State	& $v$	&${\Lambda}$ &	${\Sigma}$ & $\Omega$ & Ma/Ca & $\tilde{E}$ (\cm)\\
\hline
1294 & 5689.048000 & 20 & 4.5 &  0.010000 & 1.7350E-02 & + & e &  V2Sigma+ & 2 & 0 &  0.5 & 0.5 & Ma & 5690.100404 \\
1295 & 6231.011000 & 20 & 4.5 & 0.010000 & 1.3244E-02 & + & e &  W2Pi & 2 & 1 & 0.5 & 1.5 & Ma & 6232.617123 \\
1296 & 6851.103324 & 20 & 4.5 & 0.943406 & 1.4816E-02 & + & e &  X2Delta & 3 & 2 &  -0.5 & 1.5 & Ca & 6851.103324  \\
1297 & 7202.193271 & 20 & 4.5 & 0.597656 & 1.2959E-02 & + & e & W2Pi+ & 2 & 1 &  -0.5 & 0.5 & Ca & 7202.193271 \\
1298 & 7364.337000 & 20 & 4.5 & 0.010000 & 1.0825E-02 & + & e & X2Delta & 4 & 2 &  0.5 & 2.5 & Ma & 7365.297752 \\
1299 & 7371.380844 & 20 & 4.5 & 0.846990 & 1.1695E-02 & + & e & V2Sigma+ & 3 & 0 & 0.5 & 0.5 & Ca & 7371.380844 \\
1300 & 7873.281254 & 20 & 4.5 & 0.880736 & 1.0294E-02 & + & e &     W2Pi & 3 & 1 & 0.5 & 1.5 & Ca & 7873.281254 \\
1301 & 8572.090783 & 20 & 4.5 & 1.248474 & 1.1323E-02 & + & e &  X2Delta & 4 & 2 &  -0.5 & 1.5 & Ca & 8572.090783 \\
\hline
\hline
\end{tabular}}
\mbox{}\\
{
$i$:   state counting number,
$\tilde{E}$: state energy term values in \cm, experimental or calculated (\textsc{Duo}),
$g_i$:  total statistical weight, equal to ${g_{\rm ns}(2J + 1)}$,
$J$: total angular momentum,
unc: uncertainty, \cm,
$\tau$: lifetime (s$^{-1}$), 
$+/-$:   total parity, e/f: rotationless parity,
State: electronic state,
$v$:   state vibrational quantum number,
$\Lambda$:  projection of the electronic angular momentum,
$\Sigma$:   projection of the electronic spin,
$\Omega$:   projection of the total angular momentum, $\Omega=\Lambda+\Sigma$, Label: 'Ma' is for \lq\lq MARVELised\rq\rq\ and 'Ca' is for Calculated, $\tilde{E}$: State energy term values in \cm, Calculated (\Duo).
}
\end{table*}

\begin{table}
\centering
\caption{Extract from the transitions file of the line list for  $^{58}$NiH.}
\tt
\label{t:trans}
\centering
\begin{tabular}{rrrrr} \hline\hline
\multicolumn{1}{c}{$f$}	&	\multicolumn{1}{c}{$i$}	& \multicolumn{1}{c}{$A_{fi}$ (s$^{-1}$)}	& \multicolumn{1}{c}{$\tilde{\nu}_{fi}$} \\  \hline
    86 & 2 & 2.5678E+00 & 3509.190241 \\
    87 & 2 & 1.6285E-02 & 3808.729012 \\
    88 & 2 & 3.4564E-01 & 5165.658904 \\
    89 & 2 & 1.0530E-02 & 5413.376723 \\
    90 & 2 & 5.9463E-02 & 6746.152635 \\
    91 & 2 & 7.5545E-03 & 6958.320256 \\
    92 & 2 & 8.9879E-03 & 8247.429047 \\
    93 & 2 & 3.8742E-03 & 8447.215900 \\
    \hline\hline
\end{tabular} \\ \vspace{2mm}
\rm
\noindent
$f$: upper  state counting number; $i$:  lower  state counting number; $A_{fi}$:  Einstein-$A$ coefficient in s$^{-1}$; $\tilde{\nu}_{fi}$: transition wavenumber in \cm.
\end{table}

\begin{figure}
\includegraphics[width=0.45\textwidth]{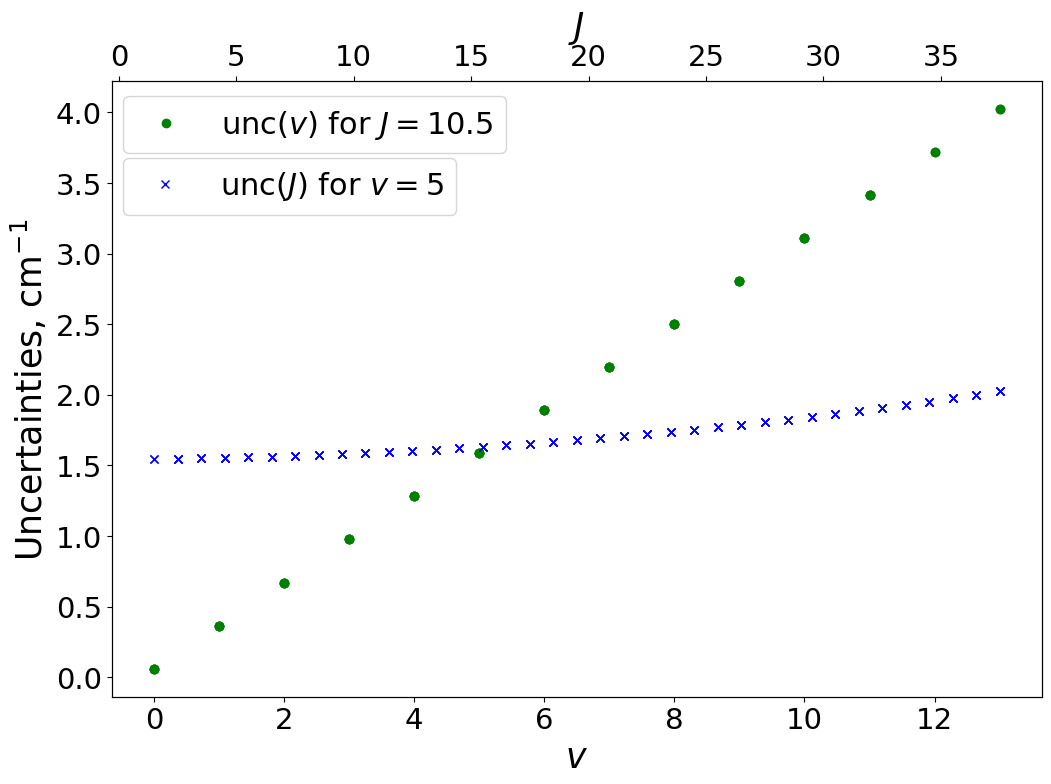}
\caption{Uncertainties as a function of $J$ (for fixed $v=5$) and as a function of $v$ (for fixed $J=10.5$) for the \X\ state of NiH.}
    \label{f:unc}
\end{figure}

We replaced calculated energy levels of $^{58}$NiH, $^{60}$NiH, and $^{62}$NiH  with the empirical ones of \citet{21HaBoPa.NiH} and the energy values of $^{58}$NiD with the empirical values of \citet{18AbShCr.NiH,19RoCrAd.NiH}, see the \texttt{Ma/Ca} column in Table \ref{t:states}). Experimental uncertainties for these  ``MARVELised'' energy term values were used in the uncertainty column of the States file (see column 5 in the table \ref{t:states}). For calculated energy term values, we assumed uncertainties to be linearly proportional to vibrational quantum numbers ($\propto v$) and quadratic proportional rotational quantum numbers ($\propto J(J+1)$). Thus, the following equation was constructed:
\begin{equation}
    \mathrm{unc}=a v + b J (J+1) +c,
    \label{eq:unc}
\end{equation}
where $a$ and $b$ are some constants of proportionality and $c$ is some constant offset (can be physically interpreted as a minimal possible uncertainty). These coefficients were separately identified for each electronic state. To do this, we looked at the distribution of Obs$-$Calc of energy term values against $J$ and $v$ separately. We chose the data points with the biggest Obs$-$Calc values and tried to adjust $a$, $b$, and $c$ in a such way that uncertainties calculated using Eq.~\eqref{eq:unc} were greater or equal than the worst Obs$-$Calc for corresponding $J$ or $v$. Same uncertainties were used for $^{58}$NiH, $^{60}$NiH, and $^{62}$NiH isotopologues, as they had nearly identical spectroscopic models.
Uncertainties were calculated independently for $^{58}$NiD isotopologue due to larger rms errors for  $^{58}$NiD.
Table \ref{t:unc} gives a summary of the values of coefficients used for uncertainties calculation. Figure \ref{f:unc} shows examples of uncertainties as a function of $J$ and $v$ for the \X\ state of NiH.

\begin{table}
    \centering
    \caption{Parameters for uncertainties estimation using Eq.~\eqref{eq:unc}.}
      \label{t:unc}
    \begin{tabular}{lccc}
    \hline
    \hline
         Isotopologue and state & $a$ & $b$ & $c$ \\
         \hline
         NiH, \X & 0.305 & 0.00033 & 0.020  \\
         NiH, \W & 0.283 & 0.00034 & 0.023  \\
         NiH, \V & 0.275 & 0.00040 & 0.013  \\
         NiD, \X & 0.389 & 0.00048 & 0.020 \\
         NiD, \W & 0.449 & 0.00055 & 0.020 \\
         NiD, \V & 0.412 & 0.00046 & 0.023 \\
    \hline
    \hline
    \end{tabular}
\end{table}


\section{NiH spectra}

\begin{figure}
\includegraphics[width=0.45\textwidth]{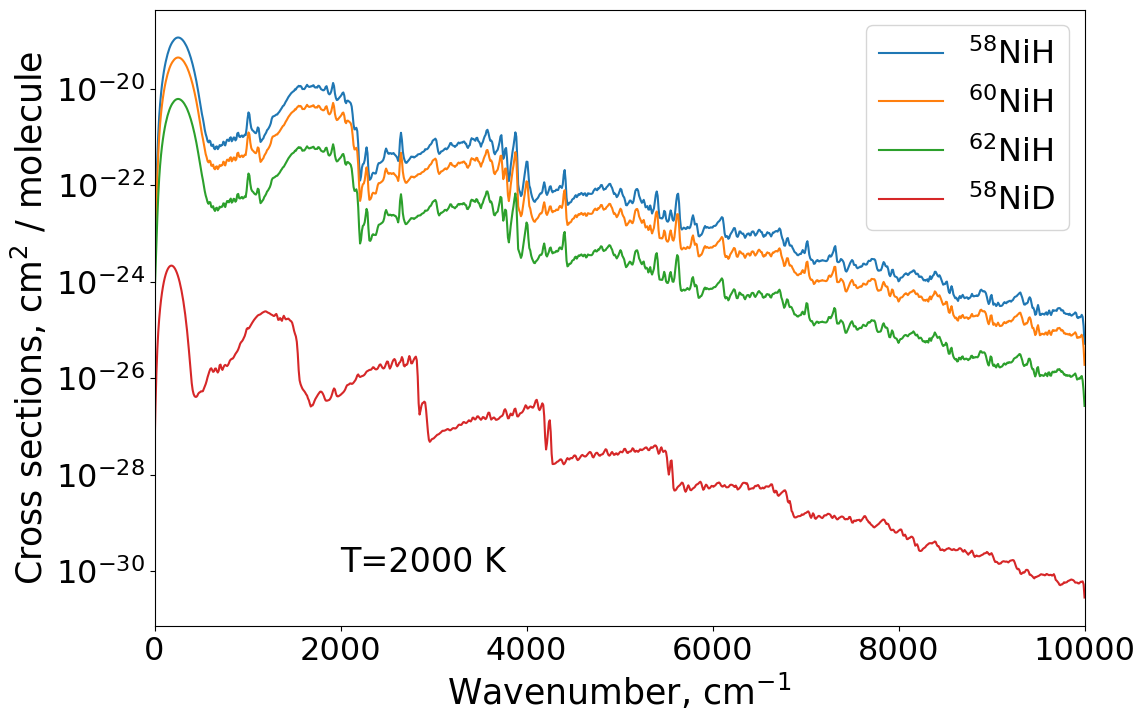}
\caption{NiH isotopologues absorption spectrum adjusted for natural abundances generated using the Gaussian profile with HWHM = 10 \cm and $T$ =  2000 K.} 
    \label{f:isotop}
\end{figure}

\begin{figure}
\includegraphics[width=0.45\textwidth]{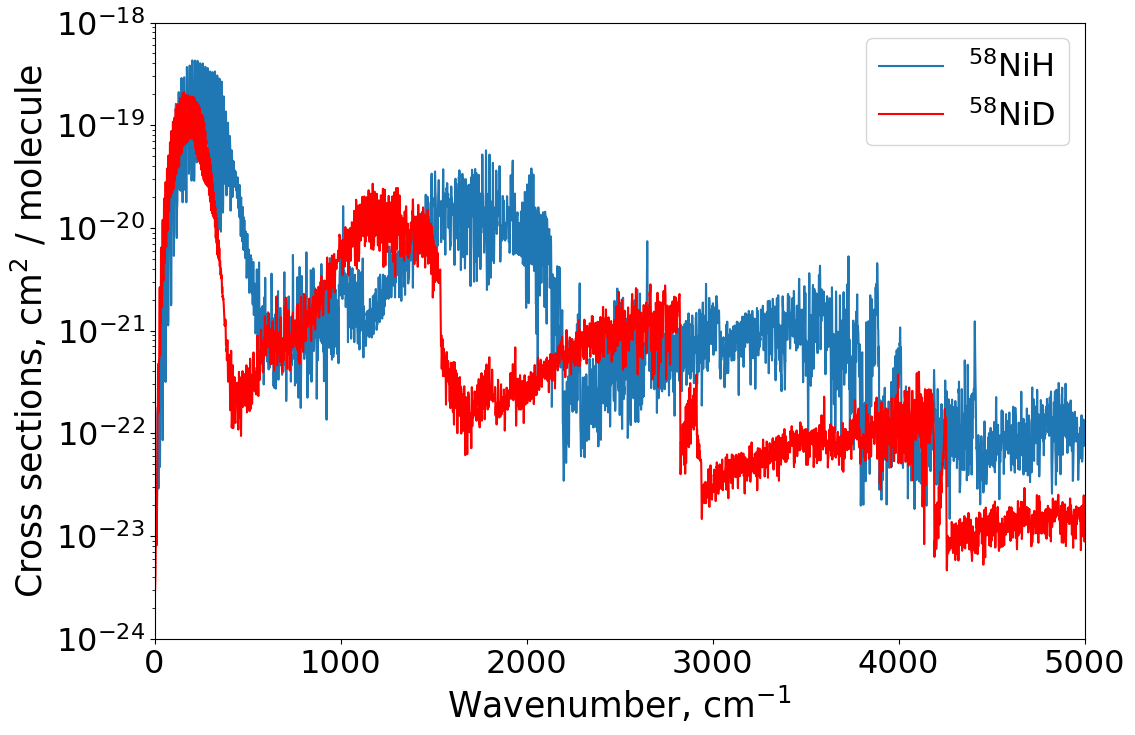}
\caption{$^{58}$NiH and $^{58}$NiD   absorption spectrum not adjusted for abundances generated using the Gaussian profile with HWHM = 1 \cm and $T$ =  2000 K.}
    \label{f:DH}
\end{figure}

\begin{figure}
\includegraphics[width=0.45\textwidth]{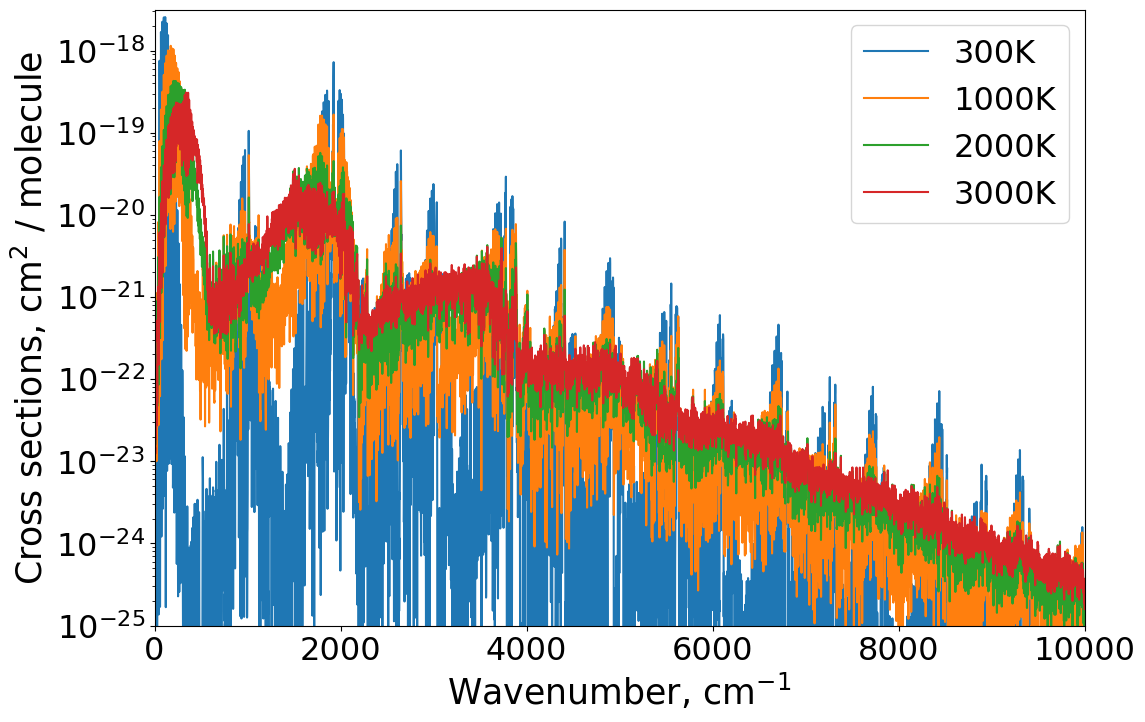}
\caption{Temperature dependence of the $^{58}$NiH absorption spectrum using the Gaussian profile with HWHM = 1 \cm.}
    \label{f:temp:58}
\end{figure}

\begin{figure}
\includegraphics[width=0.45\textwidth]{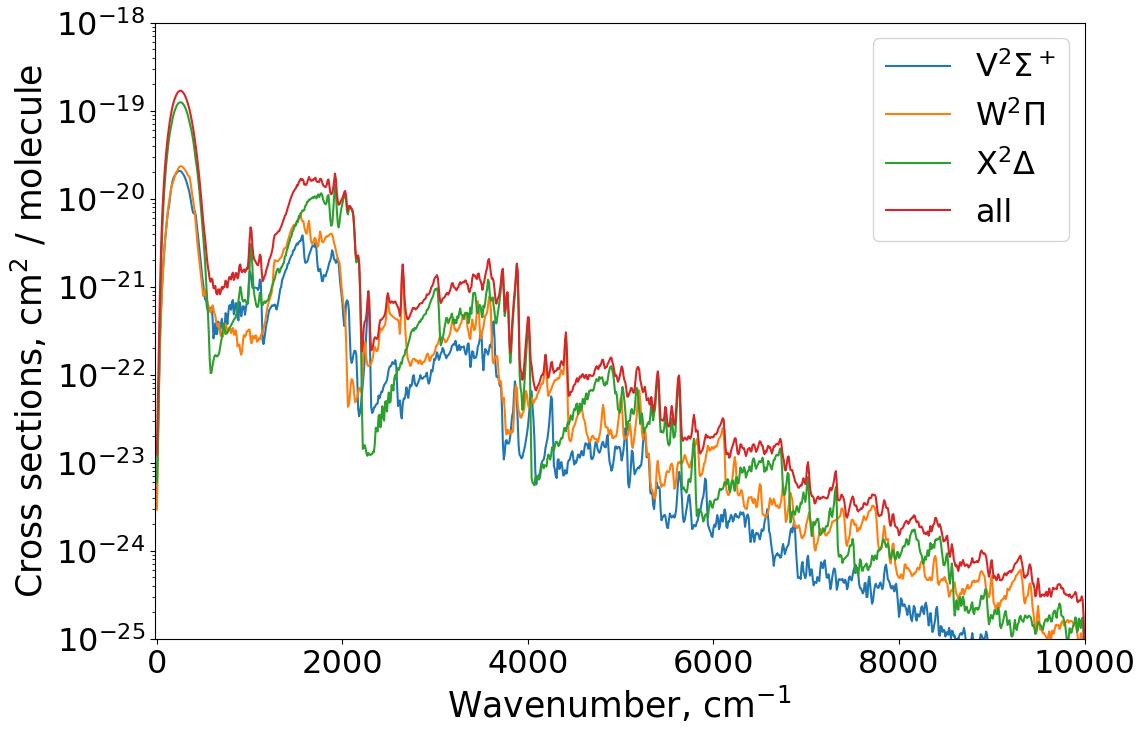}
\caption{$^{58}$NiH absorption spectrum generated using the Gaussian profile with HWHM = 10 \cm\ at $T$ =  2000 K and transitions to different upper electronic states.} 
    \label{f:state:58}
\end{figure}

Figures  \ref{f:isotop} -- \ref{f:state:58} plot four exemplary spectra of NiH. These were generated  using the code \exocross\ \citep{jt708} with settings for a simple Gaussian line profile with HWHM = 1 \cm\ or HWHM = 10 \cm. Fig.~\ref{f:isotop} shows the spectra at $T$ = 2000 K of the four isotopologues, adjusted for their abundances. Terrestrial abundances of different isotopologues of NiH were assumed to be proportional to the ratios of atomic isotopes (e.g. $^{60}$NiH to $^{58}$NiH ratio is the same as  $^{60}$Ni to $^{58}$Ni). Abundances of nickel isotopes were extracted from \citet{NNDC}. Deuterium's abundance was assumed to be similar to that on Jupiter \citep{01He.D}.

Figure \ref{f:DH} shows the spectra of $^{58}$NiH and $^{58}$NiD not adjusted for their abundances in the shorter range of $0 - 5000$ \cm. While $^{58}$NiH, $^{60}$NiH, and $^{62}$NiH have nearly identical spectra due to a small relative change in mass of nickel atom, $^{58}$NiD gives larger shifts  as the relative change in mass between hydrogen and deuterium is much larger. There is a frequency shift, which is expected due to a change in mass. Moreover, there is an intensity drop, which is caused by the partition function of NiD being greater than that of NiH (as can be seen in Fig.~\ref{f:pf:DH}).


Figure \ref{f:temp:58} shows the spectra of $^{58}$NiH at different temperatures: 300 K, 1000 K, 2000 K, and 3000 K. Figure \ref{f:state:58} shows the spectra of NiH corresponding to transitions where the upper level is specifically either in \X\ or \W\ or \V\ electronic state at $T$ = 2000 K and all of them summed up. There are no experimental or observational spectra of any of the NiH isotopologues in the range of 0--10000 \cm, so we cannot compare our generated spectra to anything. However, our spectra show the exponential decay in the intensity with the growing wavenumber, which agrees with our expectations.

\begin{figure*}
\includegraphics[width=0.45\textwidth]
{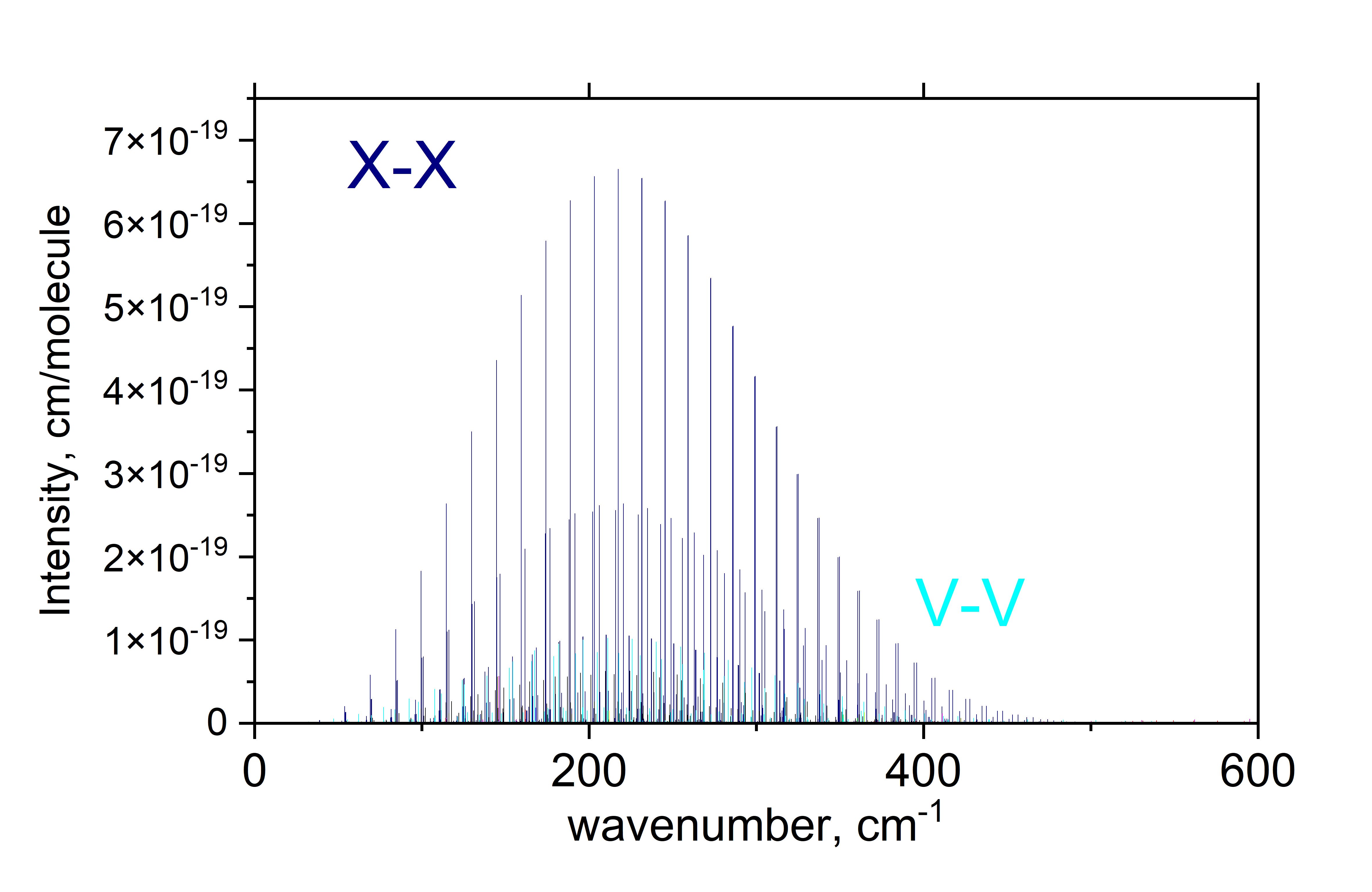}
\includegraphics[width=0.45\textwidth]{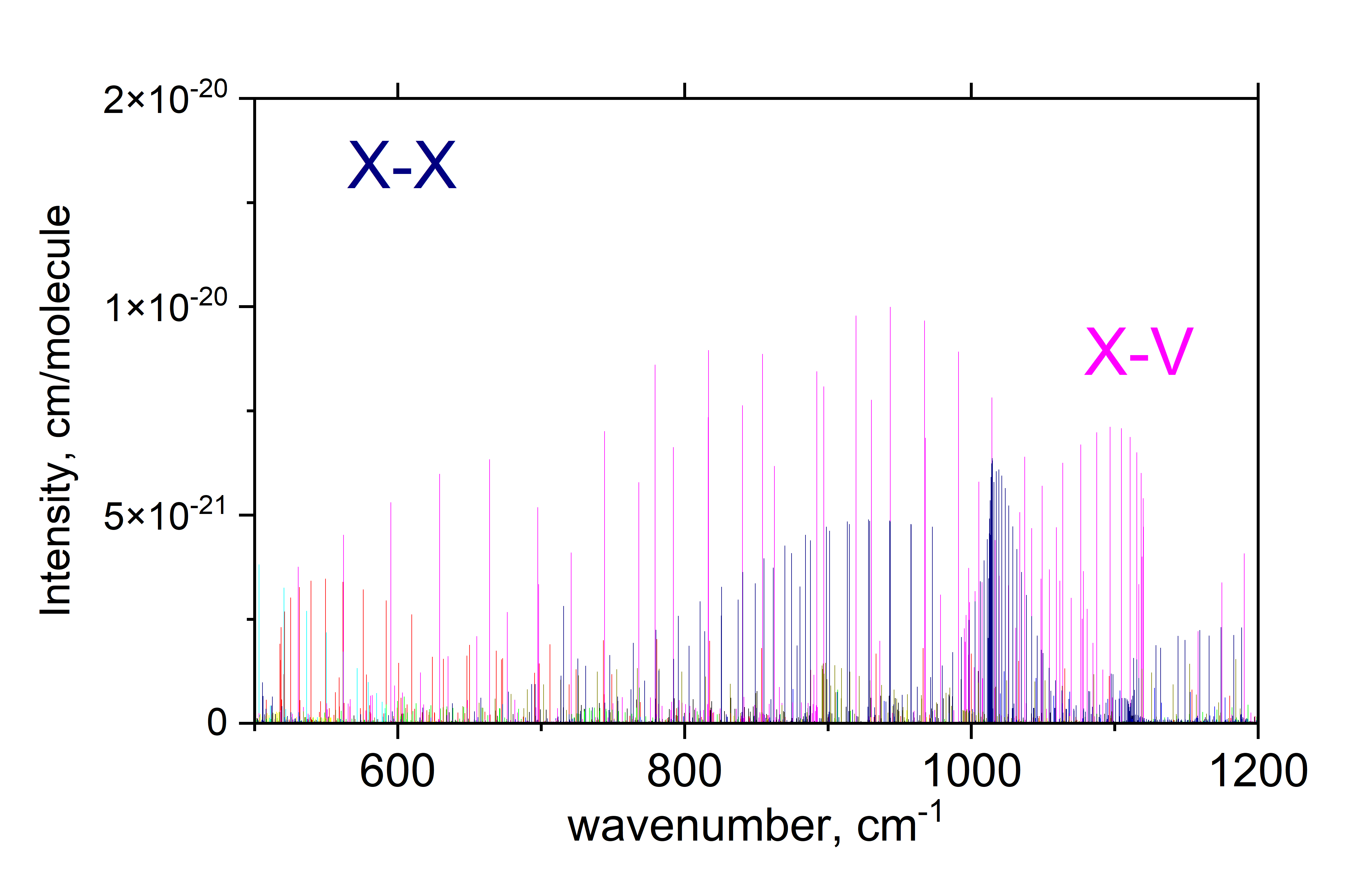}
\includegraphics[width=0.45\textwidth]{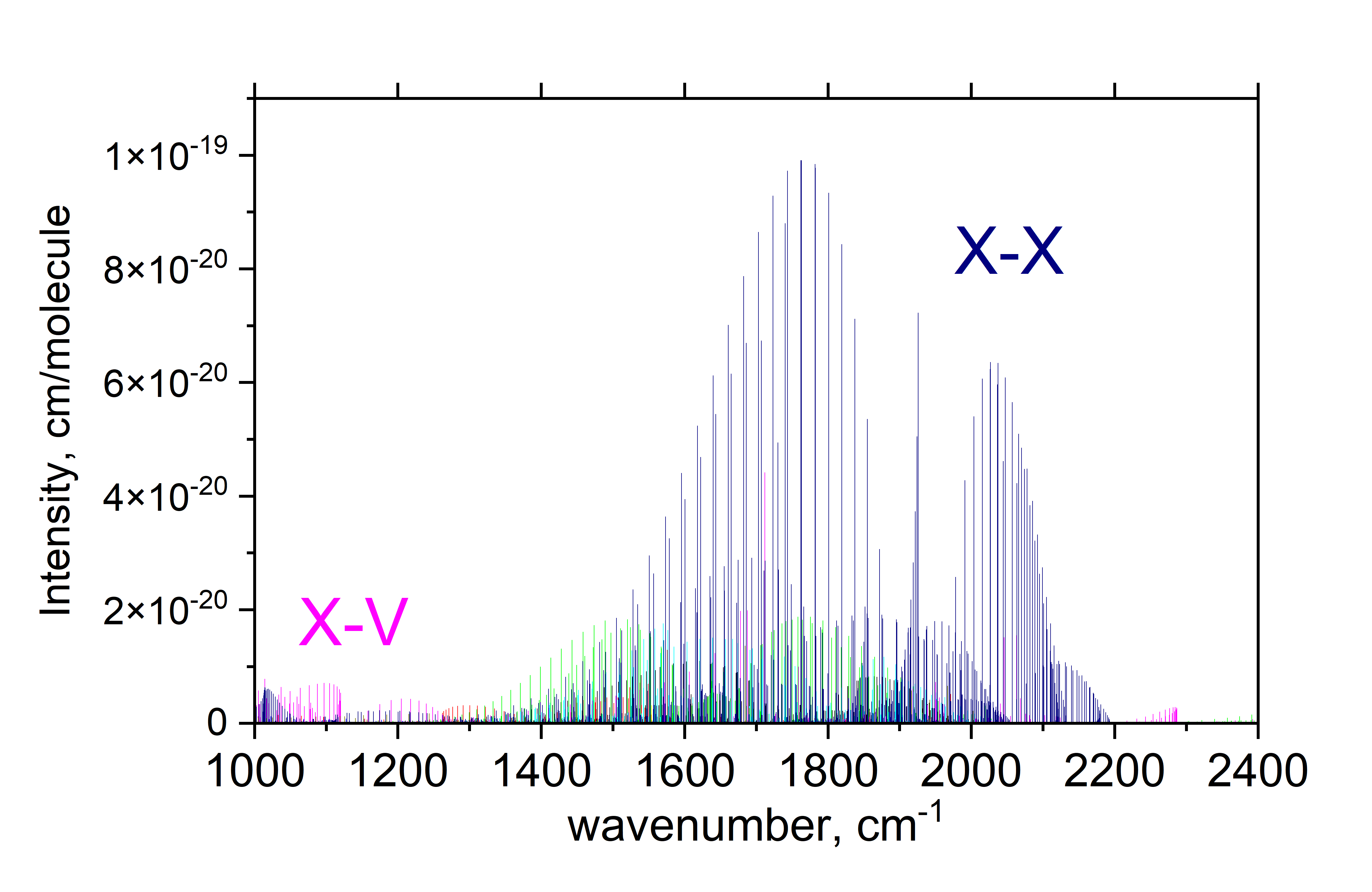}
\includegraphics[width=0.45\textwidth]{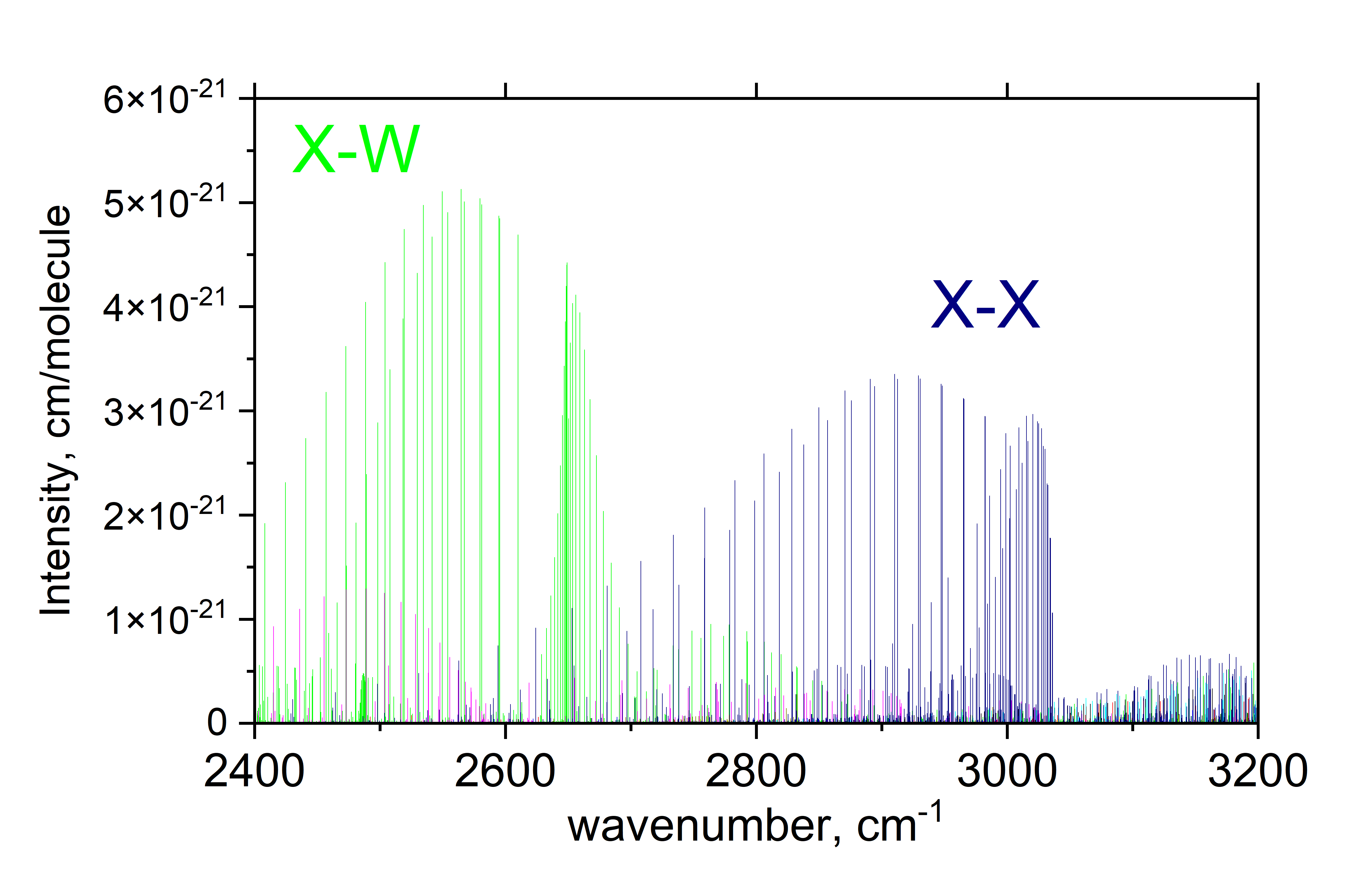}
\includegraphics[width=0.45\textwidth]{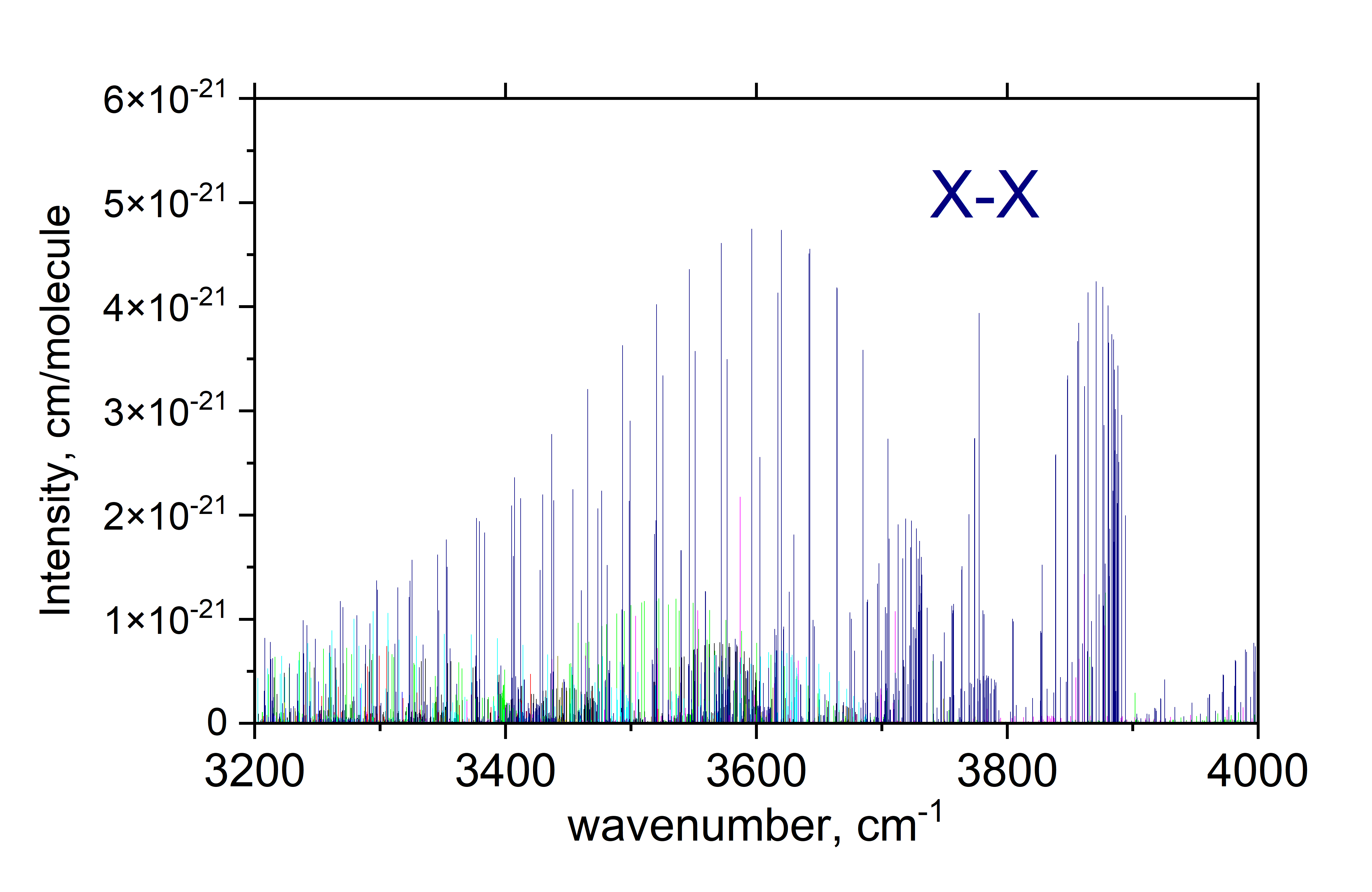}
\includegraphics[width=0.45\textwidth]{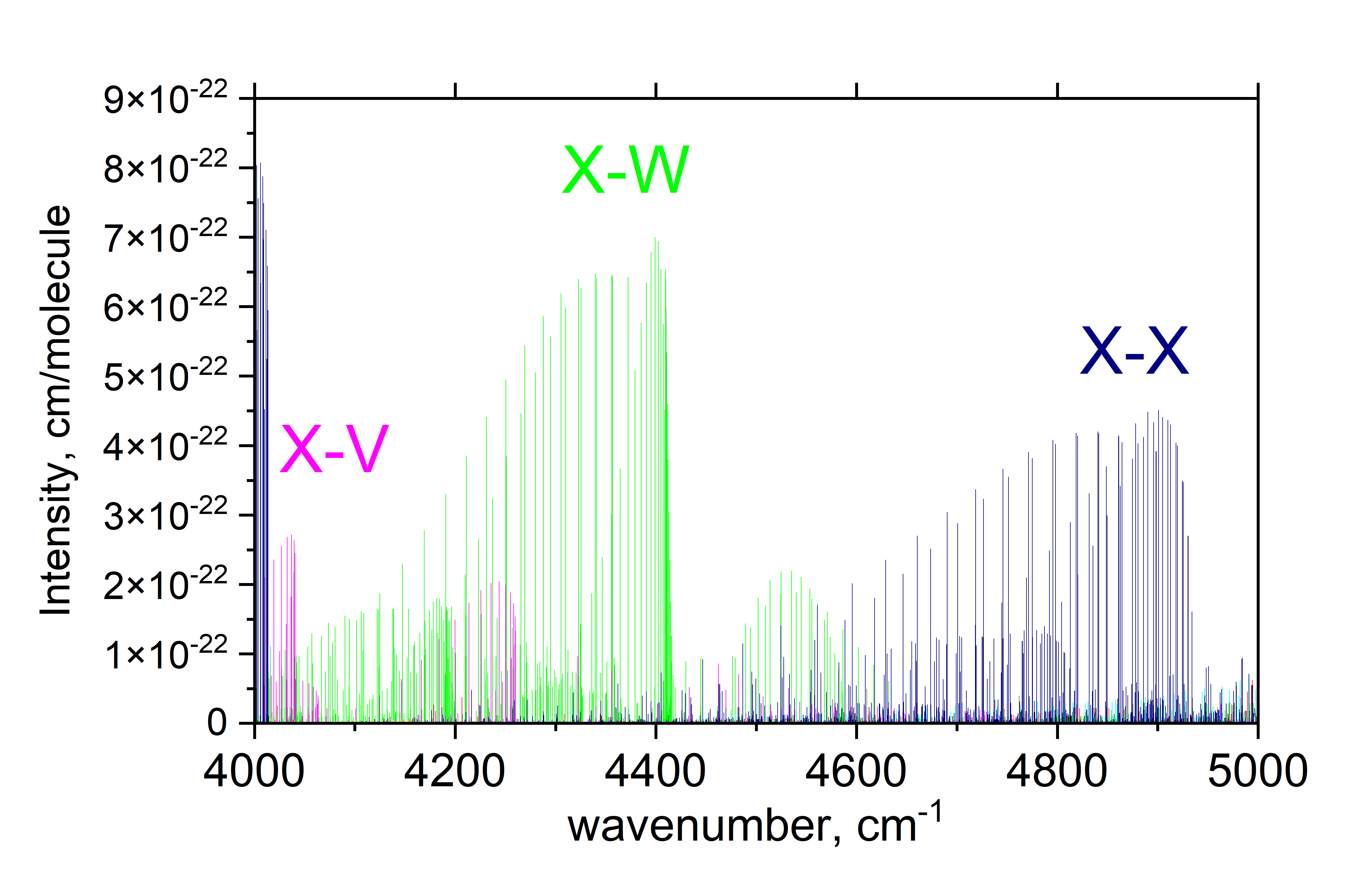}

\caption{$^{58}$NiH  absorption spectra, $T$ =  1500 K showing line intensities and divided into six spectroscopic windows below 5000~\cm. }
    \label{f:T:1500K}
\end{figure*}

In Figure \ref{f:T:1500K} a detailed illustration of the absorption bands of $^{58}$NiH is provided by showing the strongest spectroscopic features from 0 to 5000~\cm\ at $T=1500$~K. This spectrum in dominated by the transitions from the ground electronic states with some presence of the \X--\W\ and \X-\V\ bands.

\section{Conclusions}

The IR line lists \name\ for four isotopologues of NiH are provided for the first time in the range up to 10000 \cm, considering transitions and energies from the three lowest electronic states \X, \W\ and \V. We recommend using these line lists for temperatures up to 3000 K. They are available online from \url{www.exomol.com}.

More experimental data are required to improve the precision of the spectroscopic model of the three states considered here, as we did not have any experimental values for levels with $J$ higher than 16.5.
More importantly, \ai\ calculations of higher electronic states (e.g. $^2 \Sigma^-$, $^4 \Phi$ and others)  are exceedingly required, especially considering the importance of NiH in the visible in astrophysical applications. The optical region has been studied experimentally, while the theory is far behind.



\section*{Acknowledgements}

This work was supported by the European Research Council (ERC) under the European Union’s Horizon 2020 research and innovation programme through Advance Grant number 883830 and the STFC Projects No. ST/Y001508/1. The authors acknowledge the use of DiRAc.  KB acknowledges the support of the UCL MAPS Faculty Research Internships scheme. The authors acknowledge the use of the DiRAC Data Intensive service (CSD3) at the University of Cambridge, managed by the University of Cambridge University Information Services  and the DiRAC Data Intensive service (DIaL2) at the University of Leicester, managed by the University of Leicester Research Computing Service on behalf of the STFC DiRAC HPC Facility (\url{www.dirac.ac.uk}). The DiRAC services at Cambridge and  Leicester were funded by BEIS, UKRI and STFC capital funding and STFC operations grants.

\section*{Data Availability}

The states, transition and partitions function files for NiH \name\ line lists can be downloaded from \href{https://exomol.com}{www.exomol.com}.  The open access programs \Duo, \textsc{ExoCross} and \textsc{pyExoCross} are available from \href{https://github.com/exomol}{github.com/exomol}.

\section*{Supporting Information}

Supplementary data are available at MNRAS online. This includes the spectroscopic models in the form of the \Duo\ input file, containing all the curves, and parameters.




\begin{thebibliography}{}
\makeatletter
\relax
\def\mn@urlcharsother{\let\do\@makeother \do\$\do\&\do\#\do\^\do\_\do\%\do\~}
\def\mn@doi{\begingroup\mn@urlcharsother \@ifnextchar [ {\mn@doi@}
  {\mn@doi@[]}}
\def\mn@doi@[#1]#2{\def\@tempa{#1}\ifx\@tempa\@empty \href
  {http://dx.doi.org/#2} {doi:#2}\else \href {http://dx.doi.org/#2} {#1}\fi
  \endgroup}
\def\mn@eprint#1#2{\mn@eprint@#1:#2::\@nil}
\def\mn@eprint@arXiv#1{\href {http://arxiv.org/abs/#1} {{\tt arXiv:#1}}}
\def\mn@eprint@dblp#1{\href {http://dblp.uni-trier.de/rec/bibtex/#1.xml}
  {dblp:#1}}
\def\mn@eprint@#1:#2:#3:#4\@nil{\def\@tempa {#1}\def\@tempb {#2}\def\@tempc
  {#3}\ifx \@tempc \@empty \let \@tempc \@tempb \let \@tempb \@tempa \fi \ifx
  \@tempb \@empty \def\@tempb {arXiv}\fi \@ifundefined
  {mn@eprint@\@tempb}{\@tempb:\@tempc}{\expandafter \expandafter \csname
  mn@eprint@\@tempb\endcsname \expandafter{\@tempc}}}

\bibitem[\protect\citeauthoryear{Abbasi, Shayesteh, Crozet  \& Ross}{Abbasi
  et~al.}{2018}]{18AbShCr.NiH}
Abbasi M.,  Shayesteh A.,  Crozet P.,   Ross A.~J.,  2018, \mn@doi [J. Mol.
  Spectrosc.] {10.1016/j.jms.2018.03.007}, 349, 49

\bibitem[\protect\citeauthoryear{Bachem, Urban  \& Nelis}{Bachem
  et~al.}{1991}]{91BaUrNe.NiH}
Bachem E.,  Urban W.,   Nelis T.,  1991, \mn@doi [Mol. Phys.]
  {10.1080/00268979100101741}, 73, 1031

\bibitem[\protect\citeauthoryear{Bagus \& Bjorkman}{Bagus \&
  Bjorkman}{1981}]{81BaBjxx.NiH}
Bagus P.~S.,  Bjorkman C.,  1981, \mn@doi [Phys. Rev. A]
  {10.1103/PhysRevA.23.461}, 23, 461

\bibitem[\protect\citeauthoryear{Barklem \& Collet}{Barklem \&
  Collet}{2016}]{16BaCoxx.partfunc}
Barklem P.~S.,  Collet R.,  2016, \mn@doi [A\&A] {10.1051/0004-6361/201526961},
  588, A96

\bibitem[\protect\citeauthoryear{Bauschlicher, Langhoff  \&
  Komornicki}{Bauschlicher et~al.}{1990}]{90BaLaKo.NiH}
Bauschlicher C.~W.,  Langhoff S.~R.,   Komornicki A.,  1990, \mn@doi [Theor.
  Chim. Acta.] {10.1007/BF01116550}, 77, 263

\bibitem[\protect\citeauthoryear{Beaton, Evenson, Nelis  \& Brown}{Beaton
  et~al.}{1988}]{88BeEvNe.NiH}
Beaton S.~P.,  Evenson K.~M.,  Nelis T.,   Brown J.~M.,  1988, \mn@doi [J.
  Chem. Phys.] {10.1063/1.454781}, 89, 4446

\bibitem[\protect\citeauthoryear{Blomberg, Siegbahn  \& Roos}{Blomberg
  et~al.}{1982}]{82BlSiRo.NiH}
Blomberg M. R.~A.,  Siegbahn P. E.~M.,   Roos B.~O.,  1982, \mn@doi [Mol.
  Phys.] {10.1080/00268978200100092}, 47, 127

\bibitem[\protect\citeauthoryear{Brown, Beaton  \& Evenson}{Brown
  et~al.}{1993}]{93BrBeEv.NiH}
Brown J.~M.,  Beaton S.~P.,   Evenson K.~M.,  1993, \mn@doi [ApJ]
  {10.1086/187012}, 414, L125

\bibitem[\protect\citeauthoryear{Chen \& Steimle}{Chen \&
  Steimle}{2008}]{08ChStxx.NiH}
Chen J.,  Steimle T.~C.,  2008, \mn@doi [Chem. Phys. Lett.]
  {10.1016/j.cplett.2008.03.056}, 457, 23

\bibitem[\protect\citeauthoryear{Diaconu, Cho, Doll  \& Freeman}{Diaconu
  et~al.}{2004}]{04DiChDo.NiH}
Diaconu C.~V.,  Cho A.~E.,  Doll J.~D.,   Freeman D.~L.,  2004, \mn@doi [J.
  Chem. Phys.] {10.1063/1.1798992}, 121, 10026

\bibitem[\protect\citeauthoryear{Douketis, Scoles, Marchetti, Zen  \&
  Thakkar}{Douketis et~al.}{1982}]{82DoScMa.ai}
Douketis C.,  Scoles G.,  Marchetti S.,  Zen M.,   Thakkar A.~J.,  1982,
  \mn@doi [J. Chem. Phys.] {10.1063/1.443345}, 76, 3057

\bibitem[\protect\citeauthoryear{Goel \& Masunov}{Goel \&
  Masunov}{2008}]{08GoeMas.CrH}
Goel S.,  Masunov A.~E.,  2008, \mn@doi [J. Chem. Phys.] {10.1063/1.2996347},
  129, 214302

\bibitem[\protect\citeauthoryear{Gray, Rice  \& Field}{Gray
  et~al.}{1985}]{85GrRiFi.NiH}
Gray J.~A.,  Rice S.~F.,   Field R.~W.,  1985, \mn@doi [J. Chem. Phys.]
  {10.1063/1.448682}, 82, 4717

\bibitem[\protect\citeauthoryear{Gray, Li  \& Field}{Gray
  et~al.}{1990}]{90GrLiFi.NiH}
Gray J.~A.,  Li M.~G.,   Field R.~W.,  1990, \mn@doi [J. Chem. Phys.]
  {10.1063/1.457732}, 92, 4651

\bibitem[\protect\citeauthoryear{Gray, Li, Nelis  \& Field}{Gray
  et~al.}{1991}]{91GrLiNe.NiH}
Gray J.~A.,  Li M.~G.,  Nelis T.,   Field R.~W.,  1991, \mn@doi [J. Chem.
  Phys.] {10.1063/1.461393}, 95, 7164

\bibitem[\protect\citeauthoryear{Guse, Blint  \& Kunz}{Guse
  et~al.}{1977}]{77GuBlKu.NiH}
Guse M.~P.,  Blint R.~J.,   Kunz A.~B.,  1977, \mn@doi [Int. J. Quantum Chem.]
  {10.1002/qua.560110504}, 11, 725

\bibitem[\protect\citeauthoryear{Havalyova, Bozhinova, Pashov, Ross  \&
  Crozet}{Havalyova et~al.}{2021}]{21HaBoPa.NiH}
Havalyova I.,  Bozhinova I.,  Pashov A.,  Ross A.~J.,   Crozet P.,  {2021},
  \mn@doi [J. Quant. Spectrosc. Radiat. Transf.]
  {{10.1016/j.jqsrt.2021.107800}}, {272}, 107800

\bibitem[\protect\citeauthoryear{Hersant, Gautier  \& Hur\'{e}}{Hersant
  et~al.}{2001}]{01He.D}
Hersant F.,  Gautier D.,   Hur\'{e} J.-M.,  2001, \mn@doi [ApJ]
  {10.1086/321355}, 554, 391

\bibitem[\protect\citeauthoryear{Hill \& Field}{Hill \&
  Field}{1990}]{90HiFixx.NiH}
Hill E.~J.,  Field R.~W.,  1990, \mn@doi [J. Chem. Phys.] {10.1063/1.459593},
  93, 1

\bibitem[\protect\citeauthoryear{Huber \& Herzberg}{Huber \&
  Herzberg}{1979}]{79HuHe.book}
Huber K.~P.,  Herzberg G.,  1979, Molecular Spectra and Molecular Structure IV.
  Constants of Diatomic Molecules.
Van Nostrand Reinhold Company, New York, \mn@doi{10.1007/978-1-4757-0961-2},
  \url {https://doi.org/10.1007/978-1-4757-0961-2}

\bibitem[\protect\citeauthoryear{Kadavathu, Lofgren  \& Scullman}{Kadavathu
  et~al.}{1987}]{87KaLoSc.NiH}
Kadavathu S.~A.,  Lofgren S.,   Scullman R.,  1987, \mn@doi [Physica Scripta]
  {10.1088/0031-8949/35/3/009}, 35, 277

\bibitem[\protect\citeauthoryear{Kadavathu, Scullman, Gray, Li  \&
  Field}{Kadavathu et~al.}{1990}]{90KaScGr.NiH}
Kadavathu S.~A.,  Scullman R.,  Gray J.~A.,  Li M.~G.,   Field R.~W.,  1990,
  \mn@doi [J. Mol. Spectrosc.] {10.1016/0022-2852(90)90011-E}, 140, 126

\bibitem[\protect\citeauthoryear{{Le Roy}}{{Le Roy}}{2017}]{LEVEL}
{Le Roy} R.~J.,  2017, \mn@doi [J. Quant. Spectrosc. Radiat. Transf.]
  {10.1016/j.jqsrt.2016.05.028}, 186, 167

\bibitem[\protect\citeauthoryear{Lee, Seto, Hirao, Bernath  \& Le~Roy}{Lee
  et~al.}{1999}]{EMO}
Lee E.~G.,  Seto J.~Y.,  Hirao T.,  Bernath P.~F.,   Le~Roy R.~J.,  1999,
  \mn@doi [J. Mol. Spectrosc.] {10.1006/jmsp.1998.7789}, 194, 197

\bibitem[\protect\citeauthoryear{Li \& Field}{Li \& Field}{1989}]{89LiFiel.NiH}
Li M.~G.,  Field R.~W.,  1989, \mn@doi [J. Chem. Phys.] {10.1063/1.455897}, 90,
  2967

\bibitem[\protect\citeauthoryear{Lipus, Simon, Bachem, Nelis  \& Urban}{Lipus
  et~al.}{1989}]{89LiSiBa.NiH}
Lipus K.,  Simon U.,  Bachem E.,  Nelis T.,   Urban W.,  1989, \mn@doi [Mol.
  Phys.] {10.1080/00268978900101911}, 67, 1431

\bibitem[\protect\citeauthoryear{Lipus, Bachem  \& Urban}{Lipus
  et~al.}{1992}]{92LiBaUr.NiH}
Lipus K.,  Bachem E.,   Urban W.,  1992, \mn@doi [Mol. Phys.]
  {10.1080/00268979200100711}, 75, 945

\bibitem[\protect\citeauthoryear{Lipus, Urban, Evenson  \& Brown}{Lipus
  et~al.}{1993}]{93LiUrEv.NiH}
Lipus K.,  Urban W.,  Evenson K.~M.,   Brown J.~M.,  1993, \mn@doi [Mol. Phys.]
  {10.1080/00268979300101461}, 79, 571

\bibitem[\protect\citeauthoryear{Marian, Blomberg  \& Siegbahn}{Marian
  et~al.}{1989}]{89MaBlSi.NiH}
Marian C.~M.,  Blomberg M. R.~A.,   Siegbahn P. E.~M.,  1989, \mn@doi [J. Chem.
  Phys.] {10.1063/1.456891}, 91, 3589

\bibitem[\protect\citeauthoryear{McCarthy, Kanamori, Steimle, Li  \&
  Field}{McCarthy et~al.}{1997}]{97McKaSt.NiH}
McCarthy M.~C.,  Kanamori H.,  Steimle T.~C.,  Li M.~G.,   Field R.~W.,  1997,
  \mn@doi [J. Chem. Phys.] {10.1063/1.474792}, 107, 4179

\bibitem[\protect\citeauthoryear{Medvedev, Meshkov, Stolyarov, Ushakov  \&
  Gordon}{Medvedev et~al.}{2016}]{16MeMeSt}
Medvedev E.~S.,  Meshkov V.~V.,  Stolyarov A.~V.,  Ushakov V.~G.,   Gordon
  I.~E.,  2016, \mn@doi [J. Mol. Spectrosc.] {10.1016/j.jms.2016.06.013}, 330,
  36

\bibitem[\protect\citeauthoryear{{National Nuclear Data Center}}{{National
  Nuclear Data Center}}{2024}]{NNDC}
{National Nuclear Data Center} 2024, NuDat database, \url
  {www.nndc.bnl.gov/nudat/}

\bibitem[\protect\citeauthoryear{Nelis, Beaton, Evenson  \& Brown}{Nelis
  et~al.}{1991}]{91NeBeEv.NiH}
Nelis T.,  Beaton S.~P.,  Evenson K.~M.,   Brown J.~M.,  1991, \mn@doi [J. Mol.
  Spectrosc.] {10.1016/0022-2852(91)90402-V}, 148, 462

\bibitem[\protect\citeauthoryear{O'Brien \& O'Brien}{O'Brien \&
  O'Brien}{2005}]{05OBOB.NiH}
O'Brien L.~C.,  O'Brien J.~J.,  2005, \mn@doi [ApJ] {10.1086/427279}, 621, 554

\bibitem[\protect\citeauthoryear{Pavlenko, Yurchenko  \& Tennyson}{Pavlenko
  et~al.}{2020}]{19PaYuTe}
Pavlenko Y.~V.,  Yurchenko S.~N.,   Tennyson J.,  2020, \mn@doi [A\&A]
  {10.1051/0004-6361/201936811}, 633, A52

\bibitem[\protect\citeauthoryear{Pouamerigo, Merchan, Nebotgil, Malmqvist  \&
  Roos}{Pouamerigo et~al.}{1994}]{94PoMeNe.NiH}
Pouamerigo R.,  Merchan M.,  Nebotgil I.,  Malmqvist P.~A.,   Roos B.~O.,
  1994, \mn@doi [J. Chem. Phys.] {10.1063/1.467411}, 101, 4893

\bibitem[\protect\citeauthoryear{Prajapat, Jagoda, Lodi, Gorman, Yurchenko  \&
  Tennyson}{Prajapat et~al.}{2017}]{jt703}
Prajapat L.,  Jagoda P.,  Lodi L.,  Gorman M.~N.,  Yurchenko S.~N.,   Tennyson
  J.,  2017, \mn@doi [MNRAS] {10.1093/mnras/stx2229}, 472, 3648

\bibitem[\protect\citeauthoryear{Robert J. Le~Roy \& Li}{Robert J. Le~Roy \&
  Li}{2011}]{11LeHaTa.MLR}
Robert J. Le~Roy Carl C.~Haugen J.~T.,  Li H.,  2011, \mn@doi [Mol. Phys.]
  {10.1080/00268976.2010.527304}, 109, 435

\bibitem[\protect\citeauthoryear{Ross, Crozet, Richard, Harker, Ashworth  \&
  Tokaryk}{Ross et~al.}{2012}]{12RoCrRi.NiH}
Ross A.~J.,  Crozet P.,  Richard C.,  Harker H.,  Ashworth S.~H.,   Tokaryk
  D.~W.,  2012, \mn@doi [Mol. Phys.] {10.1080/00268976.2012.655336}, 110, 2019

\bibitem[\protect\citeauthoryear{Ross, Crozet, Adam  \& Tokaryk}{Ross
  et~al.}{2019}]{19RoCrAd.NiH}
Ross A.~J.,  Crozet P.,  Adam A.~G.,   Tokaryk D.~W.,  2019, \mn@doi [J. Mol.
  Spectrosc.] {10.1016/j.jms.2019.06.003}, 362, 45

\bibitem[\protect\citeauthoryear{Ruette, Blyholder  \& Head}{Ruette
  et~al.}{1984}]{84RuBlHe.NiH}
Ruette F.,  Blyholder G.,   Head J.,  1984, \mn@doi [J. Chem. Phys.]
  {10.1063/1.446968}, 80, 2042

\bibitem[\protect\citeauthoryear{Sauval \& Tatum}{Sauval \&
  Tatum}{1984}]{84SaTaxx.partfunc}
Sauval A.~J.,  Tatum J.~B.,  1984, \mn@doi [ApJS] {{10.1086/190980}}, 56, 193

\bibitem[\protect\citeauthoryear{Scullman, Lofgren  \& Kadavathu}{Scullman
  et~al.}{1982}]{82ScLoKa.NiH}
Scullman R.,  Lofgren S.,   Kadavathu S.~A.,  1982, \mn@doi [Physica Scripta]
  {10.1088/0031-8949/25/2/008}, 25, 295

\bibitem[\protect\citeauthoryear{Shaji, Nunn, O'Brien  \& O'Brien}{Shaji
  et~al.}{2008}]{08ShNuBr.NiH}
Shaji S.,  Nunn J.,  O'Brien J.~J.,   O'Brien L.~C.,  2008, \mn@doi [ApJ]
  {10.1086/523696}, 672, 722

\bibitem[\protect\citeauthoryear{Shaji, Song, Li, O'Brien  \& O'Brien}{Shaji
  et~al.}{2009}]{09ShSoLi.NiH}
Shaji S.,  Song A.,  Li M.,  O'Brien J.~J.,   O'Brien L.~C.,  2009, \mn@doi
  [Can. J. Phys.] {10.1139/P08-136}, 87, 583

\bibitem[\protect\citeauthoryear{Steimle, Nachman, Shirley, Fletcher  \&
  Brown}{Steimle et~al.}{1990}]{90StNaSh.NiH}
Steimle T.~C.,  Nachman D.~F.,  Shirley J.~E.,  Fletcher D.~A.,   Brown J.~M.,
  1990, \mn@doi [Mol. Phys.] {10.1080/00268979000100691}, 69, 923

\bibitem[\protect\citeauthoryear{Tennyson \& Yurchenko}{Tennyson \&
  Yurchenko}{2012}]{jt528}
Tennyson J.,  Yurchenko S.~N.,  2012, \mn@doi [MNRAS]
  {10.1111/j.1365-2966.2012.21440.x}, 425, 21

\bibitem[\protect\citeauthoryear{Tennyson, Hill  \& Yurchenko}{Tennyson
  et~al.}{2013}]{jt548}
Tennyson J.,  Hill C.,   Yurchenko S.~N.,  2013, in 6$^{th}$ international
  conference on atomic and molecular data and their applications ICAMDATA-2012.
  AIP, New York, pp 186--195, \mn@doi{10.1063/1.4815853}

\bibitem[\protect\citeauthoryear{Tennyson et~al.,}{Tennyson
  et~al.}{2020}]{jt810}
Tennyson J.,  et~al., 2020, \mn@doi [J. Quant. Spectrosc. Radiat. Transf.]
  {10.1016/j.jqsrt.2020.107228}, 255, 107228

\bibitem[\protect\citeauthoryear{Tennyson et~al.,}{Tennyson
  et~al.}{2024}]{jt939}
Tennyson J.,  et~al., 2024, \mn@doi [J. Quant. Spectrosc. Radiat. Transf.]
  {10.1016/j.jqsrt.2024.109083}, 326, 109083

\bibitem[\protect\citeauthoryear{Vallon, Richard, Crozet, Wannous  \&
  Ross}{Vallon et~al.}{2009}]{09VaRiCr.NiH}
Vallon R.,  Richard C.,  Crozet P.,  Wannous G.,   Ross A.,  2009, \mn@doi
  [ApJ] {10.1088/0004-637X/696/1/172}, 696, 172

\bibitem[\protect\citeauthoryear{Walch \& Bauschlicher}{Walch \&
  Bauschlicher}{1983}]{83WaBaxx.NiH}
Walch S.~P.,  Bauschlicher C.~W.,  1983, \mn@doi [J. Chem. Phys.]
  {10.1063/1.445301}, 78, 4597

\bibitem[\protect\citeauthoryear{Walch, Bauschlicher  \& Langhoff}{Walch
  et~al.}{1985}]{85WaBaLa.NiH}
Walch S.~P.,  Bauschlicher C.~W.,   Langhoff S.~R.,  1985, \mn@doi [J. Chem.
  Phys.] {10.1063/1.449704}, 83, 5351

\bibitem[\protect\citeauthoryear{Werner, Knowles, Knizia, Manby  \&
  Sch\"utz}{Werner et~al.}{2012}]{MOLPRO}
Werner H.-J.,  Knowles P.~J.,  Knizia G.,  Manby F.~R.,   Sch\"utz M.,  2012,
  \mn@doi [WIREs Comput. Mol. Sci.] {10.1002/wcms.82}, 2, 242

\bibitem[\protect\citeauthoryear{Yurchenko, Lodi, Tennyson  \&
  Stolyarov}{Yurchenko et~al.}{2016}]{Duo}
Yurchenko S.~N.,  Lodi L.,  Tennyson J.,   Stolyarov A.~V.,  2016, \mn@doi
  [Comput. Phys. Commun.] {10.1016/j.cpc.2015.12.021}, 202, 262

\bibitem[\protect\citeauthoryear{Yurchenko, Sinden, Lodi, Hill, Gorman  \&
  Tennyson}{Yurchenko et~al.}{2018a}]{jt711}
Yurchenko S.~N.,  Sinden F.,  Lodi L.,  Hill C.,  Gorman M.~N.,   Tennyson J.,
  2018a, \mn@doi [MNRAS] {10.1093/mnras/stx2738}, 473, 5324

\bibitem[\protect\citeauthoryear{{Yurchenko}, {Al-Refaie}  \&
  {Tennyson}}{{Yurchenko} et~al.}{2018b}]{ExoCross}
{Yurchenko} S.~N.,  {Al-Refaie} A.~F.,   {Tennyson} J.,  2018b, \mn@doi [A\&A]
  {10.1051/0004-6361/201732531}, 614, A131

\bibitem[\protect\citeauthoryear{Yurchenko, Al-Refaie  \& Tennyson}{Yurchenko
  et~al.}{2018c}]{jt708}
Yurchenko S.~N.,  Al-Refaie A.~F.,   Tennyson J.,  2018c, \mn@doi [A\&A]
  {10.1051/0004-6361/201732531}, 614, A131

\bibitem[\protect\citeauthoryear{Zou \& Liu}{Zou \& Liu}{2007}]{07ZoLixx.NiH}
Zou W.,  Liu W.,  2007, \mn@doi [J. Comput. Chem.] {10.1002/jcc.20742}, 28,
  2286

\makeatother
\end{thebibliography}





\bsp	
\label{lastpage}
\end{document}